# Strain-voltage and current-voltage Scanning Probe Microscopy (SPM) response of ionic semiconductor thin films: probing of deformation potential

A.N. Morozovska,[1,1] E.A. Eliseev,[2] S.L. Bravina[3] and S.V. Kalinin[4,2]

[1] Institute of Semiconductor Physics, National Academy of Science of Ukraine,
41, pr. Nauki, 03028 Kiev, Ukraine

[2] Institute for Problems of Materials Science, National Academy of Science of Ukraine,
3, Krjijanovskogo, 03142 Kiev, Ukraine

[3] Institute of Physics, National Academy of Science of Ukraine,
46, pr. Nauki, 03028 Kiev, Ukraine

[4] The Center for Nanophase Materials Sciences, Oak Ridge National Laboratory,
Oak Ridge, TN 37922

## Abstract

We performed analytical calculations of the current-voltage and strain-voltage response of the heterostructure like "charged SPM tip electrode / gap / ionic-semiconductor film" caused by the local changes of

(a) ions concentration (*stoichiometry contribution*);

(b) acceptors (donors) charge state (*recharging contribution via ionic radius variation*);

(c) free electrons (holes) concentration (*electron-phonon coupling via the deformation potential*).

The contribution (b) into the strain-voltage SPM was not calculated previously, while the contribution (c) was not even predicted before, while our estimations performed for correlated oxides show that strength of (c) appeared comparable with (a,b).

For the case of ion-blocking tip and substrate electrode mainly the changes in holes (electrons) concentration contribute into the voltage-dependent mechanical displacement of the film surface, directly registered by strain SPM. Thus, we predict that the SPM measurements of the ionic semiconductor surface displacement could provide important information about the local changes of the acceptors (donors) charge state and electron (hole)-phonon correlations via the deformation potential.

We evolve analytical formalism capable to describe the current-voltage and strain-voltage response frequency spectra in the films of correlated oxides like $p$-La$_{1-x}$Sr$_x$MnO$_{3-\delta}$ and La$_{1-}$ $_x$Sr$_x$CoO$_{3-\delta}$, where the oxygen vacancies are mobile acceptors and inherent Jahn-Teller distortions are responsible for strong electron-phonon coupling.

---

[1] morozo@i.com.ua

[2] sergei2@ornl.gov

We obtained the great variety of the nonlinear static and dynamic current-voltage and strain-voltage hysteresis loops in the ionic semiconductor thin film with mobile acceptors (donors) and holes (electrons). Some types of the current-voltage hysteresis with pronounced memory window and double loops are observed experimentally in the correlated oxides and resistive switching materials like $p$-La$_{1-x}$Sr$_x$MnO$_{3-\delta}$ and La$_{1-x}$Sr$_x$CoO$_{3-\delta}$, while predicted strain-voltage hysteresis of piezoelectric-like and butterfly-like shape requires experimental justification by SPM.

## 1. Introduction

The Scanning Probe Microscopy (SPM) methods, based on force or current detection, appeared promising tools to study the physical phenomena in the ion-containing semiconductor materials used in the energy storage and memory devices. Actually, Kalinin et al have proposed that local electrochemical dynamics in ionic semiconductors can be studied using voltage-strain coupling [1, 2, 3]. In this method, the periodically biased conductive SPM tip concentrates electric field in a small volume of the material, resulting in redistribution of mobile ions through diffusion and electromigration mechanisms. The associated changes in molar volume and strains results into periodic surface displacement detected by an SPM tip. This method may be called as Electrochemical Strain Microscopy (ESM), similarly to the well-known Piezoresponse Force Microscopy [4, 5, 6, 7]. The frequency dependent diffusion strain response of the one-dimensional electrochemically active system to periodic electric bias was analyzed in Ref.[8]. Then the image formation mechanism in ESM for the case of a single-step purely diffusion process, derive the *local strain-voltage responses* in frequency and time domains, and analyze the sensitivity and resolution limits was analyzed [9].

### *1.1. Carriers electromigration and diffusion in ionic semiconductors: pitfalls of the current theories*

The comprehensive analytical theory of charge carriers electromigration and diffusion in the ionic semiconductors and their thin films still represents a challenging task, since the current-voltage response of the materials were analyzed mainly numerically and only in the framework of the Boltzmann approximation for chemical potential and/or Debye linear screening theory assuming constant conductivity. The approximations and assumptions are invalid in the regions of space charge accumulation, which typically appear near the surfaces and interfaces of the ionic semiconductor film. Some authors even neglect the electromigration of ions as well as the acceptors (donors) mobility and bandwidth.

For instance, Svoboda and Fischer [10] considered the internal stress relaxation in thin films due to the vacancies diffusion only, Tangera et al analyzed [11] the distribution of one type space charge in oxide film between blocking electrodes, but the current was regarded absent. Cheng and Verbrugge [12] considered the lithium diffusion in spherical particle taking into account diffusion-induced stresses, but neglecting charges accumulation and electromigration. Gil et al [13, 14] analyzed current-voltage characteristics of metal/semiconductor film/metal structures assuming small variations of holes (electrons) and mobile acceptors (donors) concentrations, valid the analytical solution were derived in linear Boltzmann approximation. Using boundary conditions involving the discharge rate for conductance currents at the interfaces, as proposed by Chang and Jaffe [15], Macdonald [16] considered mobile electrons and holes, while supposing the charged ions uniformly distributed independently on applied voltage, supposed small in comparison with thermal energy. Chen [17] compared two approximate models (local electro neutrality and constant electric field) with numerical solution of Nernst-Plank-Boltzmann equations for fluxes of electrons and oxygen vacancies. Jamnik and Maier [18] proposed equivalent circuit for the model system with constant ionic conductivity. Franceschetti and Macdonald [19] considered exact solution of the Nernst-Plank equation for steady state of the system with holes, electrons and immobile charged defects. Also they numerically simulated transient currents as system response to step changes of applied bias. The theoretical background for different titration methods was proposed by Weppner et al [20] using linear diffusion model without electromigration.

All these papers [10-20] used the Boltzmann approximation for chemical potential and/or Debye linear screening theory assuming constant conductivity. Moreover, either volt-ampere or capacitance-voltage characteristics are considered for different semiconductor systems, while the effects of *strain generation* during the ions electromigration and/or recharging are usually not considered. The latter effect is crucial for the performance of the modern ions-containing energy storage materials and memory devices. However in the most of papers devoted to this problem (see e.g. [21], [22]) the space charge layers developed during the ions diffusion and emerging electric fields are usually ignored, since most efforts are made to consider the non-planar geometry, which makes the problem very cumbersome.

*1.2. Local strains originated from the non-stoichiometry and ions recharging in correlated oxides*

Well-known effect of the stoichiometry on the local strain is the linear dependence of lattice constants on the composition of solid solution (Vegard law of chemical expansion). Recent experimental studies of correlated oxides reported about additional contributions to chemical expansion (besides *non-stoichiometry*), related to either oxygen vacancies initiation or more generally *ions recharging*. Some examples for correlated oxides like $La_{1-x}Sr_xCo_yFe_{1-y}O_{3-\delta}$, $La_{0.6}Sr_{0.4}Co_{0.2}Fe_{0.8}O_{3-\delta}$, $La_{1-x}Sr_xCoO_{3-\delta}$, $Sr(Fe_xTi_{1-x})O_{3-\delta}$ are listed below.

Adler [23] analyzed the temperature and oxidation-state dependence of lattice volume in $La_{1-x}Sr_xCo_yFe_{1-y}O_{3-\delta}$ ceramics in terms of thermal and chemical expansion. Similar effect of lattice expansion due to the oxygen non-stoichiometry was observed earlier by the different authors (see e.g. Refs.[24, 25, 26]). Bishop et al [27] studied the chemical expansion and oxygen non-stoichiometry of undoped and Gd-doped cerium oxide exposed to different partial pressures of oxygen and found that the contribution to a chemical expansion could be attributed to the larger crystal radius of cerium $Ce^{3+}$ compared to the cerium $Ce^{4+}$. Phenomenological models accounting for the difference in the dopant cation radius and charge as well as the formation of oxygen vacancies have been used to explain experimental results for fluorite-structure oxides [28, 29] assuming linear relations resembling Vegard law. Lankhorst et al.[30] have found that the oxygen chemical potential in $La_{1-x}Sr_xCoO_{3-\delta}$ decreases almost linearly with the electron occupation number increase. They interpreted the observed behavior as the change of the Fermi level upon gradually filling up states in a broad electron band with electrons induced by vacancy formation and Sr doping. This model is broadly used to relate the partial oxygen pressure and the vacancy concentration, i.e. the non – stoichiometry [31, 32].

In many cases the migration of the defect (vacancy or ion) results in the reduction of surrounding affecting the chemical expansion. On the other hand, it is not always possible to distinguish experimentally between the lattice expansion induced by the vacancy/ion migration and the changes of the defect atomic radius. The latter should be derived from the independent experiments, e.g. the concentration of vacancies remains constant, but their charge state is controlled by the injection of free carriers.

*1.3. Local strains originated from the electron-phonon coupling associated with Jahn-Teller distortion in correlated oxides*

Strong *electron-phonon coupling* associated with the *local Jahn-Teller distortion* was proposed as a possible origin of this very unusual behaviour of materials with transition-metal ions [33]. Coupling between orbital occupancy and the Jahn-Teller distortion can play a major role as a driving force of symmetry breaking, because the orbital occupation may strongly couple to the lattice (anion distortion) in some cases [33]. *Jahn-Teller distortions* are typical for correlated oxides like $La_{1-x}Sr_xMnO_{3-\delta}$, $La_{1-x}Sr_xCoO_{3-\delta}$ and even $SrFe_xTi_{1-x}O_{3-\delta}$.

For instance, structural studies of $SrFe_xTi_{1-x}O_{3-\delta}$ as a function of composition and iron oxidation state have been performed [34] by means of XRD, Fe and Ti K-edge XAS, vibrational Raman and infrared spectroscopy. The combination of results obtained by XAS and vibrational spectroscopy strongly supports the presence of the Jahn-Teller distortion around $Fe^{4+}$ ions, most pronounced for composition $x=0.03$ and decreasing for higher iron concentrations. The decrease of the Jahn-Teller effect with increasing $x$ can be understood qualitatively by the change in the electronic structure of the materials from insulator to metal.

Mitchell et al. [35] reported about cooperative Jahn-Teller distortion when studied the structural properties of $La_{1-x}Sr_xMnO_{3-\delta}$ using neutron powder diffraction as a function of both Sr doping ($0<x<0.225$). In the orthorhombic phase ($x>0.125$), $MnO_6$ octahedra are irregular, and the rhombohedral-to-orthorhombic phase transition can be understood as a cooperative Jahn-Teller distortion of these octahedral, as a consequence of the orbitally degenerate electronic state of the $Mn^{3+}$ ion. As the $Mn^{3+}/Mn^{4+}$ ratio increases under the oxygen pressure decrease, the electronic energy gained by removing the electronic degeneracy eventually outweighs the elastic forces opposing distortion of the octahedral network, and a cooperative distortion of the octahedral network ensues.

Moreover, the band gap of $La_{1-x}Sr_xMnO_{3-\delta}$ (~ 1 eV) is mainly determined by the Jahn-Teller distortion [36]. Since the deformation potential, originated from the *electron-phonon coupling*, is directly related with the band gap in the narrow gap semiconductors, Fermi level in (half) metals and with the charge gap in correlated metal-insulators [37, 38], local strain-voltage response of correlated oxides like p-$La_{1-x}Sr_xMnO_{3-\delta}$ could provide the important information about the local band structure and Jahn-Teller distortions. However, the theory of the ionic semiconductor strain-voltage response measured by SPM was not elaborated.

### *1.4. Motivation of the current theoretical research*

The conclusions, which could be drawn from the above literature review, are the following

(a) The current-voltage response of the ionic semiconductors and their thin films theoretically were analyzed mainly numerically and only in the framework of conventional Boltzmann approximation.

(b) Analytical results for the current-voltage spectra were obtained only in the linear drift-diffusion theory.

(c) Conception and theory of the ionic semiconductor local strain-voltage response measured by SPM were not elaborated.

These facts motivate our theoretical study. In the paper we performed analytical and numerical calculations of the current- and local strain-voltage response of the ionic semiconductor films with realistic DOS of the mobile acceptors, donors, electrons and holes, and compared the results with the ones obtained in the Boltzmann approximation.

Original part of the paper is organized as following. Section 2 contains the model description for the calculations of the electronic properties and local strain-voltage response of the heterostructure "SPM tip electrode/dielectric gap/ionic semiconductor film". Analytical dependences of the space charge vs the electrochemical potential and band structure are derived for the rectangular and stretched exponential DOS in the Section 3. In the Section we also analyze corresponding analytical expressions for the chemical potential of the strongly doped ionic semiconductor with mobile acceptors (donors). In the Section 4 the static 1D-distributions of potential and space charge are calculated and analyzed. The analytical results were obtained in the linear Debye approximation for electrostatic potential $\varphi$ and up to the cubic nonlinearity $\varphi^3$, when the Debye approximation becomes inappropriate with the increase of the SPM tip voltage amplitude. SPM current-voltage response, its frequency spectra, depth distributions of the potential, electric field space charge and current are analyzed in the Section 5. Fermi quasi-levels, space charge and current distributions are considered in the Subsection 5.1. Dynamic current-voltage response is calculated analitically within in the linear drift-diffusion theory in the Subsection 5.2. Nonlinearity effect on the dynamic current-voltage response is analyzed in the Subsection 5.3. Analytical results of the SPM strain-voltage response calculations are analyzed in the Section 6. The static limit of the nonlinear strain-voltage response is analyzed in the Subsection 6.1. The linear finite-frequency response is calculated in the Subsection 6.2. Nonlinearity effect on the dynamic strain-voltage response is analyzed in the Subsection 6.3, where we demonstrated that only the changes in holes (electrons) concentrations contribute into the total surface displacement of the ionic film if the tip and substrate electrode are chosen acceptor blocking. These sections are followed by the brief discussion and summary remarks [Section 7].

## 2. Model for the calculations of the electronic properties and local electromechanical response of the heterostructure "SPM tip / gap / ionic semiconductor film"

Geometry of the considered asymmetric heterostructure "SPM tip / gap / ionic semiconductor film / substrate electrode" is shown in **Fig. 1a.** Electric potential $U$ is applied to the squashed tip of the SPM probe, substrate electrode is earthed. The semiconductor film is regarded thick enough to have a continuous band structure [**Fig. 1b**]. Free electrons in the conductive band (*n*) and holes (*p*) in the valence band are considered. Energy $E_g$ is the band gap width, μ is the chemical potential level, which is the Fermi level at zero temperature [**Fig. 1c**].

In the correlated oxides like $La_{1-x}Sr_xMnO_3$ and $La_{1-x}Sr_xCoO_3$ *oxygen atoms* and *vacancies* are acceptors with stoichiometry concentration $N_a$, which can create the energy quasi-band with the intrinsic halfwidth $\delta E_a$, corresponding density of states (DOS) $g_a(\varepsilon)$ and activation energy range $\{E_a - \delta E_a, E_a + \delta E_a\}$. Cations like La, Sr, Mn, Co etc can be donors with stoichiometry (or doping) concentration $N_d$, which can create the energy quasi-band with the intrinsic width $\delta E_d$, DOS $g_d(\varepsilon)$ and activation energy range $\{E_d - \delta E_d, E_d + \delta E_d\}$. The acceptors (donors) are allowed to be neutral or singly ionized. The neutral acceptors (donors) are *immobile*, only the charged ones could be *mobile* [14].

Approximated DOS are shown in **Fig. 1c** by empty well-localized distributions of widths $\delta E_m$ (hereinafter subscript $m = a, d, n, p$ denotes acceptors, donors, electrons and holes). All energies $E_m$ are counted from the bottom of the conductive band. Under the validity of the Nernst-Einstein relation, $\frac{\eta_m}{D_m} = \frac{e}{k_B T}$, the lower estimation of the acceptors (donors) quasi-band is defined by the inequality $\delta E_m \geq \frac{eD_m}{\eta_m} \sim k_B T$ ($D_m$ is the diffusion coefficient and $\eta_m$ is the carriers mobility). Denoted the band gap width as $E_g$, we define the sequence of bands as $-E_g < -E_a < -E_d < 0$, while typically the chemical potential level $\mu < 0$, but its position should be determined self-consistently

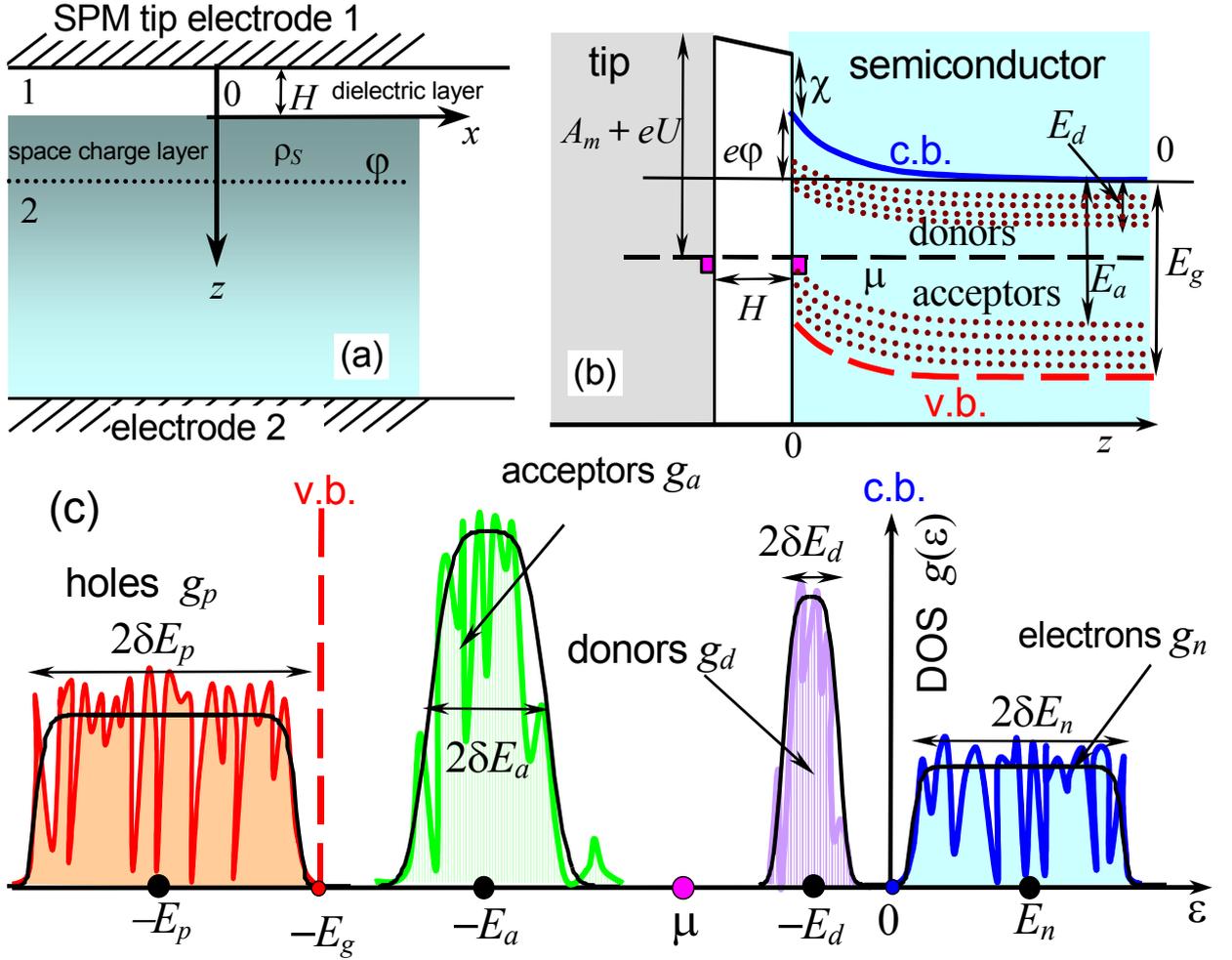

**Fig. 1.** (Color online). (a) Geometry of the considered asymmetric heterostructure "conducting SPM tip / gap / ionic semiconductor film / substrate electrode". (b) Schematic bend structure at z=0: $A_m$ is the work function difference, $U$ is the voltage difference applied to the tip electrode at $z = -H$, $\varphi$ is the electric potential, $\chi$ is the electron affinity in semiconductor, $\mu$ is the chemical potential level. (c) DOS for the holes, electrons, donors and acceptors. Filled regions with irregular boundaries schematically show the realistic DOS. Stretched exponential approximations are shown by black solid curves.

Vacuum or air dielectric or gap of thickness $H$ between the charged SPM tip electrode and the ionic semiconductor film is regarded thick enough to prevent noticeable tunneling current between the tip and the film surface, but no Zener breakdown appears. Sometimes its thickness is hardly to control. When the system is in the thermodynamic equilibrium the generalized fluxes are absent ($J = 0$), since the chemical potential $\mu$ is constant inside the semiconductor film and $J \sim -\nabla\mu = 0$. For the case the electrochemical potential $\zeta(z) = \mu + e\varphi(z)$ determines the equilibrium space charge concentration in dependence on the distance z from the interface:

(a) $n(\zeta) = \int_{-\infty}^{\infty} d\varepsilon \cdot g_n(\varepsilon) f(\varepsilon - \zeta)$ is the concentration of electrons in the conductive band;

(b) $p(\zeta) = \int_{-\infty}^{\infty} d\varepsilon \cdot g_p(\varepsilon) f(\zeta - \varepsilon)$ is the concentration of holes in the valence band;

(c) $N_d^+(\zeta) = \int_{-\infty}^{\infty} d\varepsilon \cdot g_d(\varepsilon) f(\zeta - \varepsilon)$ is the concentration of ionized donors. For *immobile* atoms, which have a delta-function DOS, $g_d(\varepsilon) = N_d \delta(E_d + \varepsilon)$, it is equal to $N_d^+(\zeta) = N_d f(E_d + \zeta)$.

(d) $N_a^-(\zeta) = \int_{-\infty}^{\infty} d\varepsilon \cdot g_a(\varepsilon) f(\varepsilon - \zeta)$ is the concentration of negatively charged acceptors. For *immobile* atoms, which have a delta-function DOS, $g_a(\varepsilon) = N_a \delta(E_a + \varepsilon)$, it is equal to $N_a^-(\zeta) = N_a f(-E_a - \zeta)$.

Hereinafter $f(x) = \dfrac{1}{1 + \exp(x/k_B T)}$ is the equilibrium Fermi-Dirac distribution function. The electrons and holes DOS includes factor 2 originated from the spin double degeneration.

In the bulk of semiconductor (or at the film earthed electrode) the electric potential vanishes ($\varphi \to 0$) and so the electroneutrality condition $n + N_a^- = N_d^+ + p$ is held even in the presence of generation-recombination processes in local equilibrium. Actually, when the system is in the steady state, the ionization of donors (acceptors) is in the local equilibrium with the electrons (holes) trapping.

Thus, the equilibrium chemical potential level µ should be found self-consistently from the integral equation $\int_{-\infty}^{\infty} d\varepsilon \big( g_p(\varepsilon) f(\mu - \varepsilon) - g_n(\varepsilon) f(\varepsilon - \mu) + g_d(\varepsilon) f(\mu - \varepsilon) - g_a(\varepsilon) f(\varepsilon - \mu) \big) = 0$. For a known DOS electroneutrality equation becomes a transcendental equation for the one number determination – chemical potential level µ. Next step is to calculate the electrostatic potential $\varphi(\mathbf{r})$ distribution from the Poisson equation for a known µ.

In order to introduce coupled equations for carrier concentration, electric potential and mechanical stress, the free energy density can be introduced explicitly as $\Omega = \sum_m \left( -Z_m c_m \varphi(\mathbf{r}) - c_m k_B T \left( \ln(c_m/c_{m0}) - 1 \right) - \beta_{jk}^{(m)} \sigma_{jk} c_m \right) - \dfrac{c_{jkli}}{2} \sigma_{jk} \sigma_{li} + \dfrac{\varepsilon_0 \varepsilon}{2} (\nabla \varphi(\mathbf{r}))^2$. Where the first term is the electrostatic energy of particles "m", with charge $Z_m$ and concentration $c_m(\mathbf{r})$, the second term is related to the entropy in the Boltzmann approximation [39], while the third term is the *generalized concentration-deformation energy*, determined by the *Vegard expansion* and *electron-phonon coupling* [37] via the *deformation tensor* $\beta_{jk}^{(m)}$ and elastic stress tensor $\sigma_{jk}(\mathbf{r})$. The first term after parenthesis is the elastic energy, $c_{jklm}$ is the tensor of elastic stiffness constants, the last term is the electrostatic energy, $\varepsilon_0$ is universal dielectric constant, $\varepsilon$ is the lattice permittivity of ionic semiconductor. The number of the deformation tensor $\beta_{jk}^{(m)}$ nontrivial components depends on the semiconductor **r**- and **k**-space symmetries.

Tensors $\beta_{ij}^{a,d}$ describe the lattice deformations caused by the small changes of the stoichiometry (*stoichiometry contribution* or *Vegard expansion* [22]) and by the changes of acceptors (donors) ionic radius, which accompany the changes of their occupation degree (*recharging contribution*) by the electrons (holes). The **r**-space symmetry group determines the Vegard expansion tensor; for isotropic or cubic media it is diagonal and reduces to scalar: $\beta_{jk}^{a,d} = \beta^{a,d}\delta_{jk}$.

The tensor $\beta_{ij}^{p,n}$ properties are determined by the **k**-space (Brillion zone) symmetry group, since the electron-phonon coupling via the deformation potential is followed by the holes or electrons spatial redistribution [38]. Also it describes the symmetric properties and the strength of the strain appeared due to the Jahn-Teller distortion [33], inherent to the correlated oxides like $p$-La$_{1-x}$Sr$_x$MnO$_3$ and $p$-La$_{1-x}$Sr$_x$CoO$_{3-\delta}$. To the best of our knowledge, $\beta_{ij}^{p,n}$ values are absent in literature for the correlated oxides, thus SPM experiments may help to determine them. Estimation of the tensors $\beta_{ij}^{p,n}$ trace for some correlated oxides materials was done in the metallic approximation; results are listed in the **Table 1**. Remarkably, that the strength of $\beta^p = \sum_i \beta_{ii}^p$ appeared comparable with $\beta^a$ for correlated oxides, while the metallic approximation significantly (up to the order of magnitude) underestimate the deformation tensor value for oxide semiconductor materials and metal-insulators with charge gap [38].

**Table 1**

| **Correlated oxide composition** | **Generalized deformation tensor** | | | | |
|---|---|---|---|---|---|
| | Ionic contribution (non stoichiometry and recharging), $\beta^a = \sum_i \beta_{ii}^a$ * | | Deformation potential of the electron-phonon coupling, $\beta^p = \sum_i \beta_{ii}^p$ ** | | |
| | $\beta^a = -\dfrac{\partial a}{a \partial c_a^{molar}}$ | $\beta^a = -\dfrac{\partial a}{a \partial c_a^{volume}}$ | $\sum_i \beta_{ii}^p = \sum_i \dfrac{-\Xi_{ii}^p}{c_{11} + 2c_{12}}$ | $\sum_i \Xi_{ii}^p \sim -\mu$ | |
| La$_{1-x}$Sr$_x$MnO$_{3-\delta}$ p-type x=0 x=0.1 x=0.2 | 0.065 [40] 0.04 [35] | 1.3 10$^{-30}$ m$^3$ 0.8 10$^{-30}$ m$^3$ | 0.65 10$^{-30}$ m$^3$ 0.54 10$^{-30}$ m$^3$ | 1.05 eV [41] 0.75 eV [42] | |

| La$_{1-x}$Sr$_x$CoO$_{3-\delta}$ p-type | From Ref. [32] | | 1.1 10$^{-30}$ m$^3$ | 1.5 eV [33] |
|---|---|---|---|---|
| x=0.2 | 0.054 + 0.511·δ | 1.1 10$^{-30}$ m$^3$ | | |
| x=0.4 | 0.055 + 0.270·δ | 1.1 10$^{-30}$ m$^3$ | | |
| x=0.7 | 0.062 + 0.187·δ | 1.2 10$^{-30}$ m$^3$ | | |
| **Comments** | *Recalculated as using stoichiometric concentration of oxygen in perovskites ~ 5 10$^{28}$ m$^{-3}$, $a$ is the lattice constant | | **Estimated from the Fermi level μ and elastic stiffness $c_{11} + 2c_{12}$=220 GPa | |

The Fermi quasi-levels could be introduced as the variational derivative of the free energy $\zeta_m(\mathbf{r}) = \pm \frac{\delta F}{\delta c_m} = \pm \left( Z_m \varphi(\mathbf{r}) + k_B T \ln(c_m(\mathbf{r})/c_{m0}) - \beta_{jk}^{(m)} \sigma_{jk} \right)$. The strain could be calculated as

$u_{jk} = -\frac{\delta F}{\delta \sigma_{jk}} = \sum_m \beta_{jk}^{(m)} c_m + c_{jkli} \sigma_{li}$. It is seen the deformation potential couples the stress field and the chemical potential, and rigorously they could not be found separately. However, in the most cases the changes of band structure due to the external pressure is rather weak (e.g., for Ge band gap changes only on about 1% for rather high strain of about 10$^{-3}$ [43]). So sort of decoupling approximation could be adopted, i.e. we neglect stress contribution, when consider the chemical potentials and carriers distribution, but we could not neglect deformation potential influence on elastic subsystem, since it is the only source of strain in our case.

In the decoupling approximation, one can recover Fermi quasi-levels $\zeta_m(\mathbf{r})$ of the electrons, ionized donors and acceptors from the dependence of their concentration $c_m$ on the electrochemical potential and calculate the conductivity currents $J = \sum_m c_m \eta_m \mathrm{grad}\zeta_m$ ($\eta_m$ is the mobility for the case of local thermal equilibrium [44]). For instance, $\zeta_n(\mathbf{r}) = -e\varphi(\mathbf{r}) + k_B T \ln(n/n_0)$ and $\zeta_p(\mathbf{r}) = -e\varphi(\mathbf{r}) - k_B T \ln(p/p_0)$ are the electron and holes Fermi quasi-levels in the Boltzmann approximation, valid in the non-degenerated case [45]. The conductivity current $J$ consist of the drift and diffusion components in the Nernst-Plank-Boltzmann approximation. The Fermi-Dirac distributions (used hereinafter) give more rigorous description of the occupation degree, which includes the Boltzmann approximation (and therefore the linear drift-diffusion approximation) for the non-degenerated case. However, the considered problem is so, that strongly doped ionic semiconductors (like correlation oxides) are typically degenerated at least with respect to the major-type carriers, making Boltzmann approximation inappropriate, especially in the regions of the space charge accumulation.

In the decoupling approximation, the *local mechanical strain*, $u_{ij}(\mathbf{r})$, is caused by the *electric voltage* applied to the tip electrode, which in turn changes the donors occupation degree $f(-\varepsilon + \zeta(\mathbf{r}))$ or/and acceptors occupation degree $f(\varepsilon - \zeta(\mathbf{r}))$ in the semiconductor film. Thus we further consider the *electromechanical strain* $u_{ij}(\mathbf{r}) \sim \sum_{m=a,d,n,p} \beta_{ij}^m \delta c_m(\mathbf{r})$. Once the "*nonstoichiometric*", "*recharging*" and "electron-phonon" contributions to the tensor $\beta_{ij}^m$ are known, in order to estimate the local strain $u_{ij}(\mathbf{r})$ one could calculate the variations

$$\delta N_d^+(\mathbf{r}) = \int_{-\infty}^{\infty} d\varepsilon \cdot g_d(\varepsilon)(f(\zeta(\mathbf{r}) - \varepsilon) - f(\mu - \varepsilon)), \qquad \delta n(\mathbf{r}) = \int_{-\infty}^{\infty} d\varepsilon \cdot g_n(\varepsilon)(f(\varepsilon - \zeta(\mathbf{r})) - f(\varepsilon - \mu)),$$

$\delta p(\mathbf{r}) = \int_{-\infty}^{\infty} d\varepsilon \cdot g_p(\varepsilon)(f(\zeta(\mathbf{r}) - \varepsilon) - f(\mu - \varepsilon))$ and $\delta N_a^-(\mathbf{r}) = \int_{-\infty}^{\infty} d\varepsilon \cdot g_a(\varepsilon)(f(\varepsilon - \zeta(\mathbf{r})) - f(\varepsilon - \mu))$. For almost immobile donor (or acceptors) we should calculate the differences $\delta N_d^+(\mathbf{r}) \approx N_d(f(E_d + \zeta(\mathbf{r})) - f(E_d + \mu))$ or $\delta N_a^-(\mathbf{r}) \approx N_a(f(-E_a - \zeta(\mathbf{r})) - f(-E_a - \mu))$ correspondingly. We further restrict the analysis to the diagonal tensor $\beta_{ij}^m = \delta_{ij} \beta_{ii}^m$ ($\delta_{ij}$ is the Kroneker delta symbol) and neglect electrostriction strains, appearing due to the internal electric field in a space charge layers (see e.g. [46, 47]).

## 3. Static dependences of the space charge on the electrochemical potential and band structure for the rectangular and stretched exponential DOS

In order to obtain analytical results, hereinafter we use the well-localized DOS, typical for strongly doped ionic semiconductors:

$$g(\varepsilon, E_m, \delta E_m) = g_m \theta(\varepsilon - E_m + \delta E_m) \theta(E_m + \delta E_m - \varepsilon), \qquad \text{(rectangular DOS)} \qquad (1a)$$

$$g(\varepsilon, E_m, \delta E_m) = g_m \exp\left(-\frac{|E_m - \varepsilon|^k}{\delta E_m^k}\right), \quad k > 0, \quad \delta E_m > 0 \quad \text{(stretched exponential DOS)} \quad (1b)$$

Here $\theta(x)$ is the unit-step function, $g_m$ is constant, subscript $m = a, d, n, p$ denotes acceptors, donors, electrons and holes. For acceptors and donors the DOS maximal value is $g_m \approx \frac{N_m}{2\delta E_m}$ for rectangular distribution (1a), $g_m = \frac{N_m}{2\delta E_m \Gamma(1 + 1/k)}$ for the stretched exponential distribution (1b) and $g_m = \frac{N_m}{2\sqrt{\pi}\delta E_m}$ for particular case of Gaussian DOS ($k = 2$). Note, that the well-localized DOS approximation is suitable for organic materials [48], narrow gap semiconductors, half-metals and metal-insulators [33].

Remarkably, that for the high exponent factors $k \gg 1$ the stretched exponential DOS (1b) tends to the rectangular one (1a) with full width $\delta E_m$ (see **Fig. 2a**). It may be very important for quantitative description of complex oxide materials, that realistic DOS could be unambiguously expanded on several well-localized stretched exponential functions with different halfwidth and amplitudes. Keeping in mind that linear analytical results admit summation, we consider only one function (1) for each type of carries for the sake of simplicity.

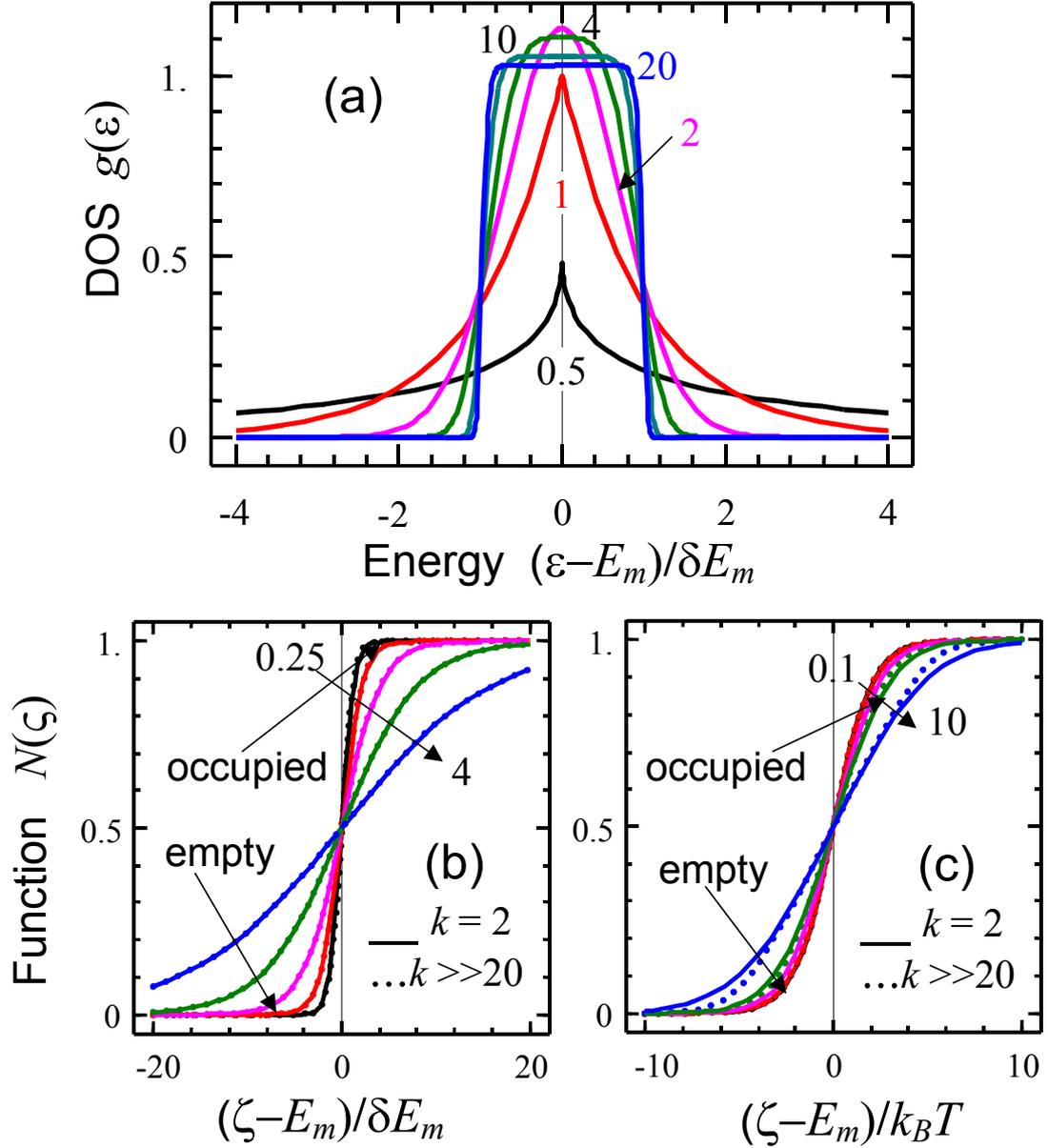

**Fig. 2.** (a) Normalized stretched exponential DOS $g(\varepsilon, E_m, \delta E_m)/N_m$ vs. the energy difference $(\varepsilon - E_m)/\delta E_m$ calculated for several values of $k = 0.5, 1, 2, 4, 10, 20$ (different curves). (b,c) Dimensionless space charge concentration $N(\zeta, E_m, \delta E_m)$ vs. $(\zeta - E_m)/\delta E_m$ calculated for $k = 2$ (solid curves), $k \to \infty$ (dotted curves) and different values of $k_B T/\delta E_m = 0.5, 1, 2, 4, 8$ (see arrows

near the curves). (c) Dimensionless space charge concentration $N(\zeta, E_m, \delta E_m)$ vs. $(\zeta - E_m)/k_B T$ calculated for $k = 2$ (solid curves), $k \to \infty$ (dotted curves) and different values of $\delta E_m/k_B T = 0.05$, 0.5, 1.5, 2.5, 5 (see arrows near the curves). Note, that $\zeta = \mu + e\varphi$ is the electrochemical potential.

Using the model DOS (1), we calculated the dependence of the carriers concentration vs. the electrochemical potential distribution:

$$n(\zeta) = g_n N(\zeta, E_n, \delta E_n), \qquad p(\varphi) = g_p P(\zeta, -E_p, \delta E_p),$$
$$N_a^-(\zeta) = g_a N(\zeta, -E_a, \delta E_a) \to N_a f(-E_a - \zeta) \quad \text{at} \quad \delta E_a \to 0, \tag{2}$$
$$N_d^+(\zeta) = g_d P(\zeta, -E_d, \delta E_d) \to N_d f(E_d + \zeta) \quad \text{at} \quad \delta E_d \to 0.$$

For the rectangular DOS given by Eq.(1a) and arbitrary temperatures the functions $N$ and $P$ were calculated as

$$N(\zeta, E, \delta E) = \int_{-\infty}^{\infty} d\varepsilon \,\theta(\varepsilon - E + \delta E)\theta(E_m + \delta E - \varepsilon) f(\varepsilon - \zeta)$$
$$= k_B T \left( \ln\left( \exp\left(\frac{\delta E}{k_B T}\right) + \exp\left(\frac{E - \zeta}{k_B T}\right) \right) - \ln\left( \exp\left(\frac{-\delta E}{k_B T}\right) + \exp\left(\frac{E - \zeta}{k_B T}\right) \right) \right) \tag{3a}$$

$$P(\zeta, E, \delta E) = \int_{-\infty}^{\infty} d\varepsilon \,\theta(\varepsilon - E + \delta E)\theta(E + \delta E - \varepsilon) f(\zeta - \varepsilon)$$
$$= k_B T \left( \ln\left( \exp\left(\frac{\delta E}{k_B T}\right) + \exp\left(\frac{\zeta - E}{k_B T}\right) \right) - \ln\left( \exp\left(\frac{-\delta E}{k_B T}\right) + \exp\left(\frac{\zeta - E}{k_B T}\right) \right) \right) \tag{3b}$$

Note, that $P(\zeta, E, \delta E) = N(-\zeta, -E, \delta E)$. The dependence of $N(\zeta, E_m, \delta E_m)$ vs. electrochemical potential $\zeta = \mu + e\varphi$ is shown in **Fig. 2b,c** by dotted curves. Both dependences are even nonlinear anti-symmetric functions of $(\zeta - E_m)$, which saturates at high values $|\zeta - E_m|/\delta E_m$ and $|\zeta - E_m|/k_B T$ correspondingly.

Using the expansion $\int_{-\infty}^{\infty} d\varepsilon (f(\varepsilon - \zeta) g(\varepsilon)) = \int_{-\infty}^{\infty} d\varepsilon (\theta(\varepsilon - \zeta) g(\varepsilon)) + \frac{\pi^2 (k_B T)^2}{6} \frac{dg(\varepsilon)}{d\varepsilon}...$, valid at not very high temperatures [see Appendix 17 in 45], the functions $N$ and $P$ should be determined for the exponential DOS given by Eq.(1b) as:

$$N(\zeta, E, \delta E) = \int_{-\infty}^{\infty} d\varepsilon \cdot f(\varepsilon - \zeta) \exp\left(-\frac{|E - \varepsilon|^k}{\delta E^k}\right) \approx$$

$$\approx \begin{cases} \dfrac{\delta E}{k}\left(\Gamma\left(\dfrac{1}{k}\right) - \Gamma\left(\dfrac{1}{k}, \dfrac{(\zeta - E)^k}{\delta E}\right)\right) - \dfrac{\pi^2 (k_B T)^2}{6} \dfrac{k(\zeta - E)^{k-1}}{\delta E^k} \exp\left(-\dfrac{(\zeta - E)^k}{\delta E^k}\right), & \zeta > E, \\ \dfrac{\delta E}{k} \Gamma\left(\dfrac{1}{k}, \dfrac{(E - \zeta)^k}{\delta E^k}\right) + \dfrac{\pi^2 (k_B T)^2}{6} \dfrac{k(E - \zeta)^{k-1}}{\delta E^k} \exp\left(-\dfrac{(E - \zeta)^k}{\delta E^k}\right), & \zeta < E. \end{cases} \quad (4a)$$

$$P(\zeta, E, \delta E) = \int_{-\infty}^{\infty} d\varepsilon \cdot f(\zeta - \varepsilon) \exp\left(-\frac{|E - \varepsilon|^k}{\delta E^k}\right) \approx$$

$$\approx \begin{cases} \dfrac{\delta E}{k}\left(\Gamma\left(\dfrac{1}{k}, \dfrac{(\zeta - E)^k}{\delta E^k}\right)\right) + \dfrac{\pi^2 (k_B T)^2}{6} \dfrac{k(\zeta - E)^{k-1}}{\delta E^k} \exp\left(-\dfrac{(\zeta - E)^k}{\delta E^k}\right), & \zeta > E, \\ \dfrac{\delta E}{k}\left(\Gamma\left(\dfrac{1}{k}\right) - \Gamma\left(\dfrac{1}{k}, \dfrac{(E - \zeta)^k}{\delta E^k}\right)\right) - \dfrac{\pi^2 (k_B T)^2}{6} \dfrac{k(E - \zeta)^{k-1}}{\delta E^k} \exp\left(-\dfrac{(E - \zeta)^k}{\delta E^k}\right), & \zeta < E. \end{cases} \quad (4b)$$

The incomplete gamma function is defined as $\Gamma(a,b) = \int_b^{\infty} dt \cdot t^{a-1} \exp(-t)$.

For the case of Gaussian DOS at $k = 2$ Eqs.(4) could be essentially simplified as:

$$N(\zeta, E, \delta E) = \frac{\sqrt{\pi}}{2} \delta E \cdot \mathrm{Erfc}\left(\frac{E - \zeta}{\delta E \sqrt{\pi}}\right) + \frac{\pi^2 (k_B T)^2}{3} \frac{E - \zeta}{\delta E^2} \exp\left(-\frac{(E - \zeta)^2}{\delta E^2}\right), \quad (5a)$$

$$P(\zeta, E, \delta E) = \frac{\sqrt{\pi}}{2} \delta E \cdot \mathrm{Erfc}\left(\frac{\zeta - E}{\delta E \sqrt{\pi}}\right) + \frac{\pi^2 (k_B T)^2}{3} \frac{\zeta - E}{\delta E^2} \exp\left(-\frac{(\zeta - E)^2}{\delta E^2}\right). \quad (5b)$$

Note, that again $P(\zeta, E, \delta E) = N(-\zeta, -E, \delta E)$ (compare with Eqs.(3)). The function $N(\zeta, E, \delta E)$ is shown in **Figs. 2b,c** by solid curves. Both dependences are even nonlinear anti-symmetric functions of $(\zeta - E_m)$, which saturates at high values $|\zeta - E_m|/\delta E_m$ and $|\zeta - E_m|/k_B T$ correspondingly.

Unexpectedly, the difference between the space charge concentration calculated by the integration with rectangular (1a) and stretched-exponential (1b) DOS with exponent factor $k \geq 2$ appeared very small (see the small distinction between solid and dotted curves in **Figs. 2b,c**). Noticeable differences appeared in the case $k < 1.5$. Additional analyses proved, that the fact originated from the smearing of the distribution details under integration. In other words, this result allows us to neglect the differences between the stretched exponential DOS with exponent factor $k \geq 2$ (e.g. between rectangular and Gaussian-like DOS) in the calculations of electric fields and elastic strain performed below and consequently it gives solid background to use the DOS that admits to obtain analytical results.

Then local electroneutrality condition $n + N_a^- = N_d^+ + p$ for the Fermi level $\mu$ determination in the bulk of semiconductor acquires the form:

$$g_n N(\mu, E_n, \delta E_n) + g_a N(\mu, -E_a, \delta E_a) = g_p P(\mu, -E_p, \delta E_p) + g_d P(\mu, -E_d, \delta E_d) \qquad (6)$$

Relatively simple analytical solution of Eq.(6) can be derived for the strongly doped semiconductor. Namely, for the case of p-type doping with $g_a = g_p$ and $g_a \gg g_n, g_d$ we obtained the following expression for the chemical potential:

$$\mu_p(E_p, \delta E_p, E_a, \delta E_a) = k_B T \ln\left(\frac{1}{2} e^{-\frac{E_p + \delta E_p}{k_B T}} \left(e^{\frac{2\delta E_a}{k_B T}} - 1\right)^{-1} \left| e^{\frac{2\delta E_a}{k_B T}} - e^{\frac{2\delta E_p}{k_B T}} \right| \right) +$$

$$+ k_B T \ln\left(-S + \sqrt{1 + 4 e^{\frac{E_p + \delta E_p - E_a + \delta E_a}{k_B T}} \left(e^{\frac{2\delta E_a}{k_B T}} - 1\right)\left(e^{\frac{2\delta E_p}{k_B T}} - 1\right)\left(e^{\frac{2\delta E_a}{k_B T}} - e^{\frac{2\delta E_p}{k_B T}}\right)^{-2}}\right) \qquad (7)$$

Where $S(\delta E_a, \delta E_p) = \text{sign}\left(e^{\frac{2\delta E_a}{k_B T}} - e^{\frac{2\delta E_p}{k_B T}}\right)$ is the sign factor. For the case of n-type doping with $g_d = g_n$ and $g_d \gg g_p, g_a$ we obtained the same function for the chemical potential, but the arguments and signs are different: $\mu_n = \mu_p(E_d, \delta E_d, -E_n, \delta E_n)$.

Dependences of the chemical potential $\mu_p$ on the acceptors activation energy $E_a$, their distribution halfwidth $\delta E_a$, temperature $T$ and conductance band halfwidth $\delta E_p$ were calculated from Eq.(7) and are shown in **Figs. 3**. The constraint $E_p - \delta E_p = E_g$ was used as shown in **Fig. 1c**.

The chemical potential $\mu_p(E_a)$ decreases linearly on $E_a$ value, and the Mott transition (i.e. $\mu_p < -E_g$) appears with $E_a$ increase as anticipated for shallow acceptors (**Fig. 3a**). The $\mu$-level decreases with $\delta E_a$ increase (compare curves 1-4 in **Fig. 3a**).

The dependence $\mu_p(\delta E_a)$ is shown in **Fig. 3b** for several values of $E_a$. The $\mu$-level decreases with $E_a$ increase (compare curves 1-4). For chosen parameters the dependences have the steep fall at small ratio $\delta E_a/E_g \ll 0.1$ and then gradually quasi-linearly decreases with further $\delta E_a/E_g$ increase.

Temperature dependence of the $\mu$-level is nonlinear, at that its increase with the temperature becomes steeper with the ratio $\delta E_a/E_g$ decrease (compare curves 1-4 in **Fig. 3c**). Potentially the Mott transition $\mu_p > 0$ is not excluded.

The dependence $\mu_p(\delta E_p)$ is shown in **Fig. 3d** for several values of $\delta E_a$. The μ-level decreases with $\delta E_p$ increase (compare curves 1-4). The region with constant positive slope and saturation are seen in the log-scale.

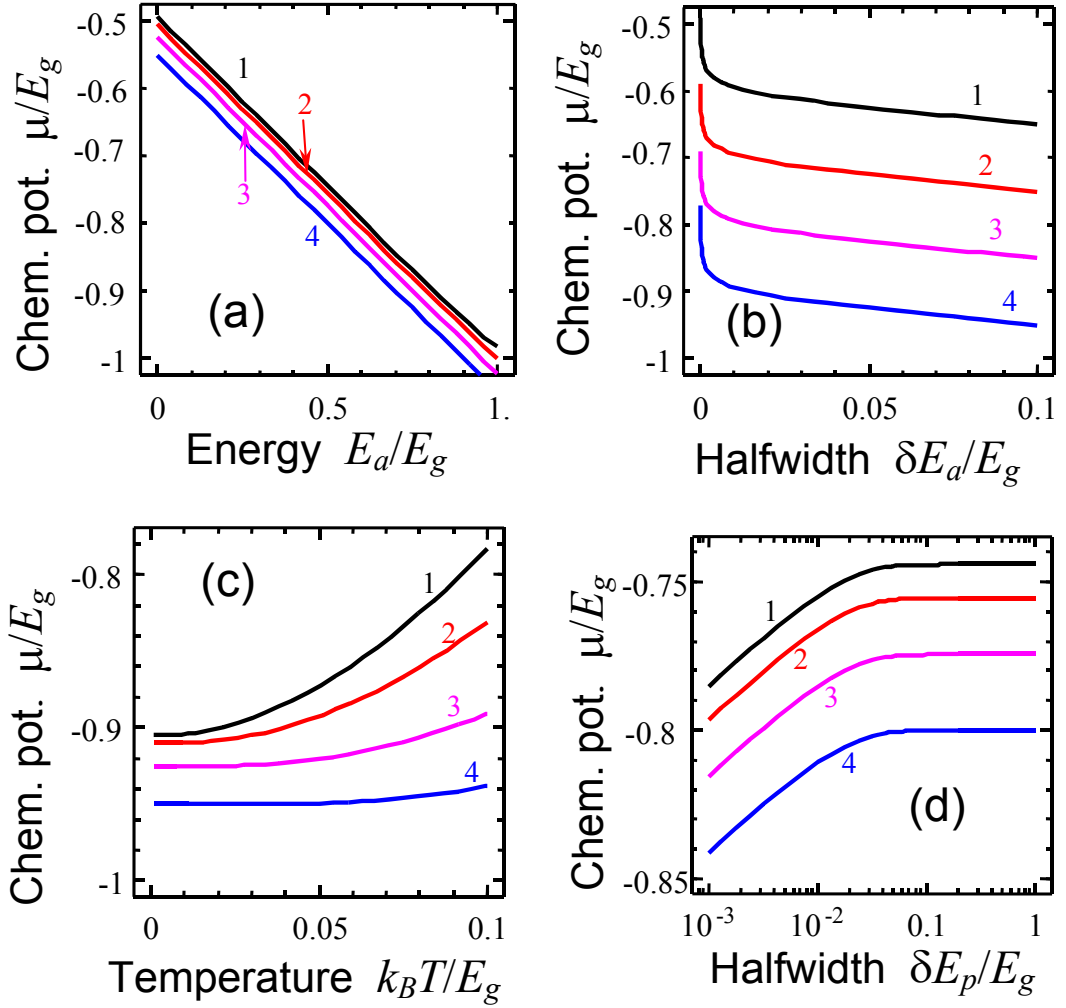

**Fig. 3.** (a) Dependence of the chemical potential $\mu_p$ normalized on the band gap width $E_g$ vs. the acceptors activation energy $E_a$ calculated at $k_B T = 0.03\,E_g$, $E_p = 1.5\,E_g$, $\delta E_p = 0.5\,E_g$ and several $\delta E_a/E_g = 0.01, 0.02, 0.05, 0.1$ (curves 1, 2, 3, 4). (b) Chemical potential $\mu_p$ vs. the acceptors distribution halfwidth $\delta E_a$ calculated at $k_B T = 0.03\,E_g$, $E_p = 1.5\,E_g$, $\delta E_p = 0.5\,E_g$ and several $\delta E_a/E_g = 0.2, 0.4, 0.6, 0.8$ (curves 1, 2, 3, 4). (c) Chemical potential $\mu_p$ vs. temperature $T$ calculated at $E_a = 0.8\,E_g$, $E_p = 1.5\,E_g$, $\delta E_p = 0.5\,E_g$ and several $\delta E_a/E_g = 0.01, 0.02, 0.05, 0.1$ (curves 1, 2, 3, 4). (d) Chemical potential $\mu_p$ vs. conductance band halfwidth $\delta E_p$ calculated at $k_B T = 0.03\,E_g$, $E_a = 0.5$ and several $\delta E_a/E_g = 0.01, 0.02, 0.05, 0.1$ (curves 1, 2, 3, 4).

To summarize the results presented in the Section 3, we derive analytical expressions for the static dependences of the space charge carriers (ionized donors, acceptors, electrons and holes) on the electrochemical potential and the band structure of the strongly doped ionic semiconductor film with rectangular or stretched exponential DOS. Unexpectedly, the difference between the dependences calculated for the rectangular and stretched-exponential DOS with exponent factor $k \geq 2$ appeared very small in the entire temperature range. The result, originated from the smearing of the DOS details under the integration, allows further neglecting the differences between the stretched exponential DOS with exponent factor $k \geq 2$ and rectangular DOS in all calculations performed below; consequently it gives a solid background to use the rectangular DOS, more suitable to obtain analytical results.

## 4. Static distributions of electric potential, field and space charge in the heterostructure SPM tip / dielectric gap / ionic semiconductor film / electrode

In order to obtain analytical results, we regard that the effective transverse size of the SPM tip, $r$, is at least several times larger than the gap thickness $H$ and the screening radius $R_S$ of the semiconductor film or, alternatively, the film thickness $h$ (see **Fig. 1a**). The assumption $r >> \min\{H + R_S, H + h\}$ allows us to solve 1D problem for electric field determination.

Poisson equations for the electrostatic potential $\varphi$ and displacement **D** for the 1D-case acquires the form:

$$\frac{d^2\varphi}{dz^2} = 0, \quad -H < z < 0 \tag{8a}$$

$$\frac{d^2\varphi}{dz^2} = -\frac{\rho_S(\varphi)}{\varepsilon_{33}^S \varepsilon_0}, \quad 0 < z < h \tag{8b}$$

The space charge density in the ionic semiconductor film has the form:

$$\rho_S(\varphi) = e\left(N_d^+(\varphi) + p(\varphi) - n(\varphi) - N_a^-(\varphi)\right). \tag{9}$$

The value of the carrier elementary charge is $e$.

Eqs.(8) should be supplemented with the boundary conditions for the electrostatic potential $\varphi(z)$:

$$\varphi(x, y, z = h) = 0, \quad \text{(earthed substrate electrode)} \tag{10}$$

$$\varphi(z = -H) = U^*, \quad \varphi(z = +0) - \varphi(z = -0) \approx U_b \quad \text{(tip electrode-dielectric gap-film)} \tag{11}$$

$$D_{2n} - D_{1n} = -\varepsilon_0\left(\varepsilon_{33}^S \frac{\partial\varphi(z=+0)}{\partial z} - \varepsilon_{33}^g \frac{\partial\varphi(z=-0)}{\partial z}\right) = \sigma_f. \quad \text{(film-dielectric gap)} \tag{12}$$

Where $U^* = \dfrac{A_m}{e} + U$, $A_m$ is the work function from the conducting tip electrode, $U$ is the voltage difference applied to the tip electrode at $z = -H$, contact built-in potential difference is $U_b = -\dfrac{\chi}{e}$ (e.g. Schottky barrier), $\varepsilon_{33}^g$ is the dielectric constant of the dielectric layer. The normal vector **n** is pointed from media 1 to media 2, the free surface charge is $\sigma_f$. Note, that the potential can be always set zero at the contact $z = h$, while the contact itself may either has contact barrier or be barrierless (ohmic).

The potential in the gap linearly depends on $z$ and has the form

$$\varphi(z) = U^* + \frac{H+z}{\varepsilon_0 \varepsilon_{33}^g}\left(\sigma_f + \varepsilon_0 \varepsilon_{33}^S \frac{\partial \varphi(+0)}{\partial z}\right), \quad -H \leq z < 0, \tag{13}$$

Substitution of Eq.(13) into the boundary condition (11-12) leads to the third kind boundary condition for the electric potential: $\varphi(+0) - H\dfrac{\varepsilon_{33}^S}{\varepsilon_{33}^g}\dfrac{\partial \varphi(+0)}{\partial z} = U^* + U_b + \dfrac{H\sigma_f}{\varepsilon_0 \varepsilon_{33}^g}$. Then allowing for Eqs.(2) electrostatic problem (8b) acquires the form

$$\begin{cases} \dfrac{d^2\varphi}{dz^2} = \dfrac{e}{\varepsilon_{33}^S \varepsilon_0}\left(\begin{array}{l} g_n N(\mu + e\varphi, E_n, \delta E_n) + g_a N(\mu + e\varphi, -E_a, \delta E_a) \\ -g_p P(\mu + e\varphi, -E_p, \delta E_p) - g_d P(\mu + e\varphi, -E_d, \delta E_d) \end{array}\right), \\ \varphi(0) - H\dfrac{\varepsilon_{33}^S}{\varepsilon_{33}^g}\dfrac{\partial \varphi(0)}{\partial z} = V, \quad \varphi(h) = 0. \end{cases} \tag{14}$$

Where the *acting voltage V* is introduced as

$$V = U^* + U_b + \frac{H\sigma_f}{\varepsilon_0 \varepsilon_{33}^g}. \tag{15}$$

Allowing for the equation (6) for the chemical potential $\mu$ the right-hand-side of Eq.(14) could be expanded into the series on potential $\varphi$ powers:

$$\rho_S(\mu + e\varphi) = e\sum_{k=1}^{\infty}\frac{(e\varphi)^k}{k!}\frac{d^k}{d\mu^k}\left(\begin{array}{l} g_n N(\mu, E_n, \delta E_n) + g_a N(\mu, -E_a, \delta E_a) \\ -g_p P(\mu, -E_p, \delta E_p) - g_d P(\mu, -E_d, \delta E_d) \end{array}\right) \tag{16}$$

In order to obtain approximate analytical solution of the nonlinear boundary problem (15) we will use the perturbation theory.

### *4.1. Debye approximation*

Under the assumption of weak field-induced bend bending Eqs.(14) can be linearized with respect to potential $\varphi$. Linearized solution was derived as

$$\varphi(z) = \varphi_0(h,V)\left(\exp\left(-\frac{z}{R_S}\right) - \exp\left(-\frac{2h-z}{R_S}\right)\right), \tag{17a}$$

$$\varphi_0(h,V) = V\left(1 - \exp\left(-\frac{2h}{R_S}\right) + \left(1 + \exp\left(-\frac{2h}{R_S}\right)\right)\frac{H}{R_S}\frac{\varepsilon_{33}^S}{\varepsilon_{33}^g}\right)^{-1}. \tag{17b}$$

The Debye screening radius $R_S$ is defined as:

$$\frac{1}{R_S^2} = \frac{e^2}{\varepsilon_{33}^S \varepsilon_0}\begin{pmatrix} 2g_n Q(\mu, E_n, \delta E_n) + g_a Q(-\mu, -E_a, \delta E_a) \\ -2g_p Q(-\mu, -E_p, \delta E_p) - g_d Q(-\mu, -E_d, \delta E_d) \end{pmatrix}. \tag{18}$$

The function $Q(\mu, E, \delta E)$ in Eq.(18) stands for the first derivative of the functions $P(\zeta, E, \delta E) = N(-\zeta, -E, \delta E)$. In particular case of rectangular DOS the function $Q(\mu, E, \delta E)$ is:

$$Q(\mu, E, \delta E) = \frac{dN(\mu, E, \delta E)}{d\mu} = \frac{dP(-\mu, -E, \delta E)}{d\mu} = \frac{e^{\frac{E+\delta E}{k_B T}} - e^{\frac{E-\delta E}{k_B T}}}{\left(1 + e^{\frac{E+\delta E-\mu}{k_B T}}\right)\left(1 + e^{\frac{E-\delta E-\mu}{k_B T}}\right)}. \tag{19}$$

Note, that in the limit $\delta E \to 0$ and $\exp\left(\frac{E-\mu}{k_B T}\right) << 1$ the function $Q(\mu, E, \delta E) \approx \frac{2\delta E}{k_B T}$ and constant $g_m = \frac{N_m}{2\delta E_m}$, so the screening radius tends to its classical limit $R_S = \sqrt{\frac{\varepsilon_{33}^S \varepsilon_0 k_B T}{e^2 N_m}}$, where $m = n$ or $p$ for the purely p-type on n-type ionic semiconductor film correspondingly [49].

### *4.2. Beyond the Debye approximation*

Note, that solution (17) may be rather rigorous for the case $|e\varphi/\mu| << 1$. Next step is to take into account the nonlinearity in Eq.(16) up to the third term, and to solve the nonlinear boundary problem (14) for the equation $\frac{d^2\varphi}{dz^2} = \frac{1}{R_S^2}(\varphi + \alpha\varphi^2 + \beta\varphi^3)$, where $\alpha$ and $\beta$ are proportional to the second and third derivatives of the space on the chemical potential. Using the expansion, the solution was derived analytically for thick films ($h >> R_S$) as:

$$\varphi(z) = \frac{4\varphi_0 \exp(-z/R_S)}{(1 - (2\alpha/3)\varphi_0 \exp(-z/R_S))^2 - 2\beta\varphi_0^2 \exp(-2z/R_S)}. \tag{20a}$$

Solution (20a) is valid at $\beta < 0$, arbitrary sign and value of the product $\alpha\varphi_0$; or at $\alpha\varphi_0 < 3/2$ and $\beta = 0$. The integration constant $\varphi_0$ should be determined from the boundary condition $\varphi(0) - H\frac{\varepsilon_{33}^S}{\varepsilon_{33}^g}\frac{\partial\varphi(0)}{\partial z} = V$ that leads to the fourth order algebraic equation for $\varphi_0$, which approximate solution was derived as:

$$\varphi_0 = \frac{3V^*}{2\left(3 + V^*\alpha + 3\sqrt{1 + (2\alpha/3)V^* + (\beta/2)(V^*)^2}\right)}, \quad V^* \equiv \frac{U^* + U_b + \left(H\sigma_f/\varepsilon_0\varepsilon_{33}^g\right)}{1 + \left(H\varepsilon_{33}^S/R_S\varepsilon_{33}^g\right)}. \quad (20b)$$

It is seen from Eq.(20b) that the solution (20a) has sense under the condition $\left(1 + \frac{2\alpha}{3}V^* + \frac{\beta}{2}(V^*)^2\right) > 0$. Note, that the introduced reduced voltage $V^*$ depends on the gap depth $H$.

Using solutions (17) or (20), z-distributions of the space charges $n(\mu + e\varphi(z))$, $p(\mu + e\varphi(z))$, $N_d^+(\mu + e\varphi(z))$ and $N_a^-(\mu + e\varphi(z))$ were calculated from nonlinear expressions (9) as the next order of perturbation theory.

Z-distributions of the electric potential and field are shown in **Figs. 4a,b** for different values of the voltage ratio $eV^*/E_g$ and sign (compare different curves) and film thickness $h \gg R_S$. It is seen from the figures that the linear Debye approximation (17) works good only for the small voltages $|eV^*| \ll E_g$. For the case $|eV^*| \geq E_g$ nonlinear solution (20) gives much more rapid vanishing of the electric potential and field into the semiconductor depth, at that the field value at the surface z=0 is much higher than in the linear Debye approximation (17).

Z-distributions of the total charge $\rho_S$, ionized acceptors, holes and electrons are shown in **Figs. 4c-f** for different sign values and of the voltage ratio $eV^*/E_g$, film thickness $h \gg R_S$. Asymmetry related with the voltage sign (compare dashed and solid curves) is mainly caused by strong p-type doping rather than rectification at the Shottky contact, since the asymmetry increases under the increase of $N_a$ with respect to $N_d$ and under the shift of acceptors level in the direction of conduction band, while the built-in difference $U_b$ is included into $V^*$ and thus shifts the curves as a whole.

Note, that the variation between the potential distributions (as well as between the field distributions), which differs only by the compensation degree $N_a/N_d = 10^4$ and $N_a/N_d = 10^{-4}$ correspondingly, are very small as anticipated from the compensation rule. Therefore we did not plot the figures for n-type semiconductors. However, for material parameters chosen in **Figs. 4** the values $R_S = 2.1$ nm and $\mu = -0.85$ eV were calculated for the ratio $N_a/N_d = 10^4$, while $R_S = 1.1$ nm, $\mu = -0.11$ eV correspond to the ratio $N_a/N_d = 10^{-4}$.

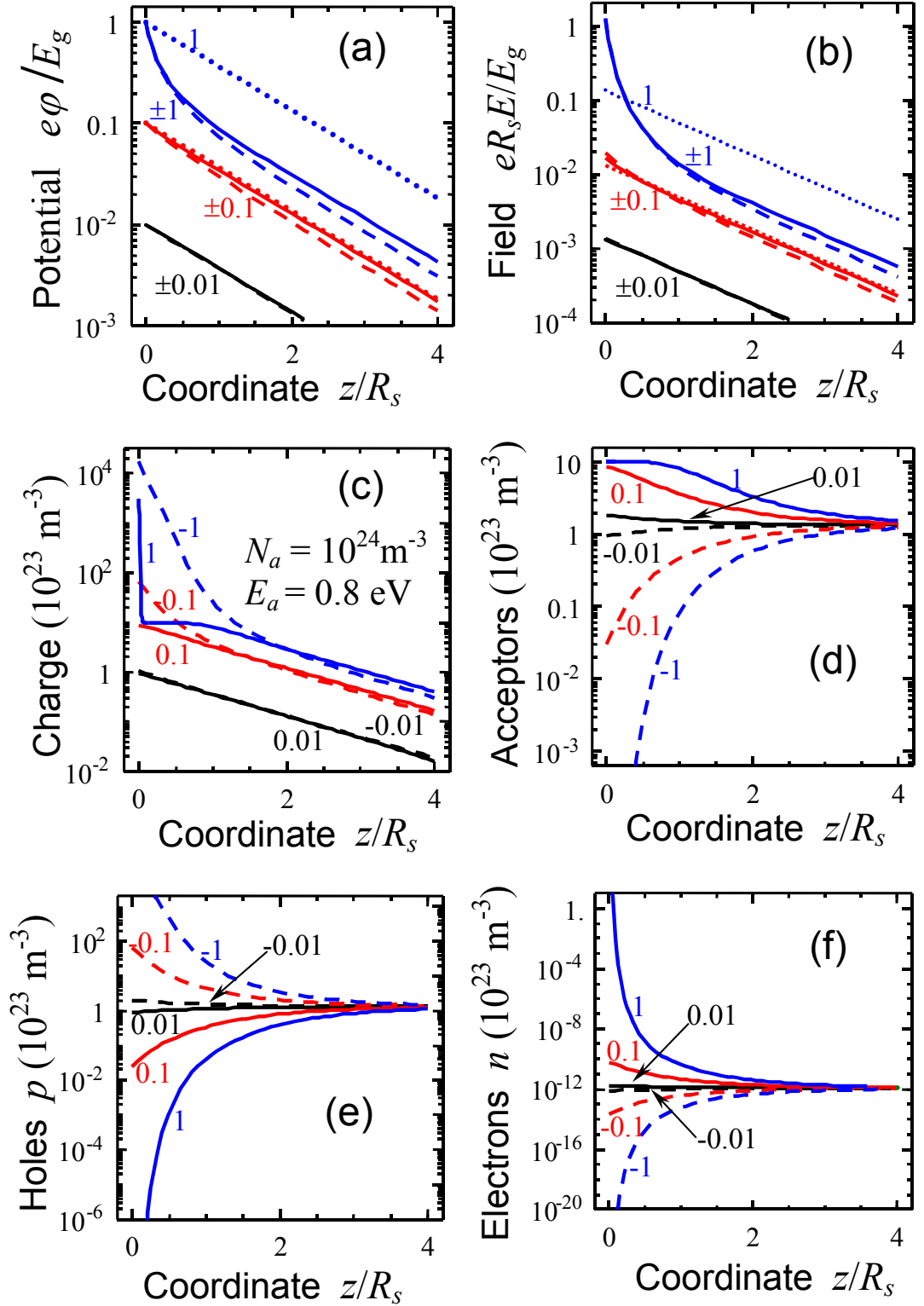

**Fig. 4.** Distributions of electric potential absolute value $|e\varphi|/E_g$ (a), and electric field absolute value $eR_S E_z/E_g$ (b), total charge (c), ionized acceptors (d), holes (e), electrons (f) calculated for different values of acting voltage $eV^*/E_g$ specified near the curves, and the following concentration of donors and acceptors atoms $N_a = 10^{24}$ m$^{-3}$ and $N_d = 10^{20}$ m$^{-3}$. Other parameters are $T = 293$ K, $E_g = 1$ eV, $\delta E_p = 0.5$ eV, $\delta E_n = 0.5$ eV, $\delta E_a = 0.026$ eV, $\delta E_d = 0.026$ eV,

$E_a$ = 0.8 eV, $E_d$= 0.1 eV, $E_n$ = 1.5 eV, $E_n$ = 0.5 eV. Dashed curves are nonlinear solution for negative voltages. Solid and dotted curves in plot (a) and (b) represent nonlinear and linear solutions given by Eqs. (20) and (17) respectively.

To summarize the results, presented in the Section 4, we derived analytical expressions for the 1D-distributions of the electrostatic potential, field and charge carriers in the heterostructure "charger SPM tip / dielectric gap / ionic semiconductor film / earthed electrode". The analytical results were analyzed in the linear Debye approximation for electrostatic potential φ and beyond the Debye approximation up to the cubic nonlinearity $\varphi^3$. We showed that the Debye approximation becomes invalid with the increase of acting voltage amplitude $V^* > k_B T/e$ and thus nonlinear effects should not be neglected for the most of realistic cases.

## 5. SPM current-voltage response of ionic semiconductor films: electric field and space charge evolution, dynamic I-V characteristics, hysteresis and memory effects

### 5.1. General problem with boundary conditions. Fermi quasi-levels. Space charge and current distributions

The total electric current is the ionic semiconductor film is $J_f = J_{Ds} + J_c$, where $J_{Ds}(z,t) = \varepsilon_0 \varepsilon_{33}^{S,g} \frac{\partial E_z}{\partial t}$ is the displacement current (also existing in the gap), and $J_c(z,t)$ is the full conductivity current existing in the semiconductor only, which should be in agreement with continuity equation $\frac{\partial \rho}{\partial t} + \frac{\partial J_c}{\partial z} = 0$. The continuity equation should be solved along with the all electrodynamics equations. For the considered 1D-case it leads to the equations: $\frac{\partial E_z}{\partial z} = \frac{\rho}{\varepsilon_0 \varepsilon_{33}^S}$ and $\frac{\partial}{\partial z}\left(\varepsilon_0 \varepsilon_{33}^S \frac{\partial E_z}{\partial t} + J_c(t)\right) = 0$.

The conductivity current $J_c(z,t) = \sum_m J_c^m$ is proportional to the gradients of the carriers Fermi quasi-levels (see Section 2 and Ref.[44]). In particular, acceptors and holes conductivity currents are $J_c^a = N_a^- \eta_a \frac{\partial \zeta_a}{\partial z}$ and $J_c^p = p \eta_p \frac{\partial \zeta_p}{\partial z}$, where $\eta_m$ is the mobility; $\zeta_a(z)$ and $\xi_p(z)$ are the acceptors and holes Fermi quasi-levels calculated as:

$$\zeta_p(z) = -e\varphi(z) - E_p + k_B T \left( \ln\left( e^{\frac{\delta E_p}{k_B T}} - e^{\frac{p(z)}{g_p k_B T} - \frac{\delta E_p}{k_B T}} \right) - \ln\left( e^{\frac{p(z)}{g_p k_B T}} - 1 \right) \right)$$

$$\approx -e\varphi(z) - E_p + k_B T \ln\left( 2\sinh\left(\frac{\delta E_p}{k_B T}\right) \right) - k_B T \ln\left( \frac{p(z)}{g_p k_B T} \right),$$

(21a)

$$\zeta_a(z) = -e\varphi(z) - E_a - k_B T \left( \ln\left( e^{\frac{\delta E_a}{k_B T}} - e^{\frac{N_a^-(z)}{g_a k_B T} - \frac{\delta E_a}{k_B T}} \right) - \ln\left( e^{\frac{N_a^-(z)}{g_a k_B T}} - 1 \right) \right)$$

$$\approx -e\varphi(z) - E_a - k_B T \ln\left( 2\sinh\left(\frac{\delta E_a}{k_B T}\right) \right) + k_B T \ln\left( \frac{N_a^-(z)}{g_a k_B T} \right).$$

(21b)

Eqs.(21) was derived directly from Eqs.(2)-(3), where we put $p = 2g_p P(\zeta_p + e\varphi, -E_p, \delta E_p)$ and $N_a^- = g_a N(\zeta_a + e\varphi, -E_a, \delta E_a)$ with $g_p = N_p/(2\delta E_p)$ and $g_a = N_a/(2\delta E_a)$. Potential distribution $\varphi(z)$ should be determined from the boundary problem (14). The approximation in Eqs.(21) corresponds to the non-degenerated statistics consistent with the Boltzmann approximation.

Concentration dependences of ionized acceptors and holes Fermi quasi-levels are shown in **Figs. 5**. It is seen from the plots, that the Boltzmann approximation (dashed lines) works well for moderate concentrations of charges $p < p_{cr}$ and $N_a^- < (N_a^-)_{cr}$. The degeneration rapidly appears with further increase of the concentrations (see vertical deviation of the solid curves from the dashed ones in **Figs. 5**). Critical concentrations, corresponding to steep increase of quasi-level value, was found from Eqs.(21) as $p_{cr} = 2g_p \delta E_p \equiv N_p$ and $(N_a^-)_{cr} = 2g_a \delta E_a \equiv N_a$. Note, that the condition $p(z) \geq p_{cr}$ and/or $N_a^-(z) \geq (N_a^-)_{cr}$ can be readily achieved in the vicinity of the semiconductor film interfaces, where the space charge accumulation takes place (see discrepancies between the solid and dashed curves in **Figs. 6**). Exactly in the regions Boltzmann approximation and consequently the linear drift-diffusion model for the case conductivity currents $J_c^a = -e\eta_a N_a^- \frac{d\varphi}{dz} + eD_a \frac{dN_a^-}{dz}$ and $J_c^p = -e\eta_p p \frac{d\varphi}{dz} - eD_p \frac{dp}{dz}$ becomes inappropriate for the description of the charge transfer. Alternatively, exact expressions (21) can be used entire the film depth.

Note, that Fermi quasi-levels for ionized donors, electrons and corresponding currents can be derived in a similar way.

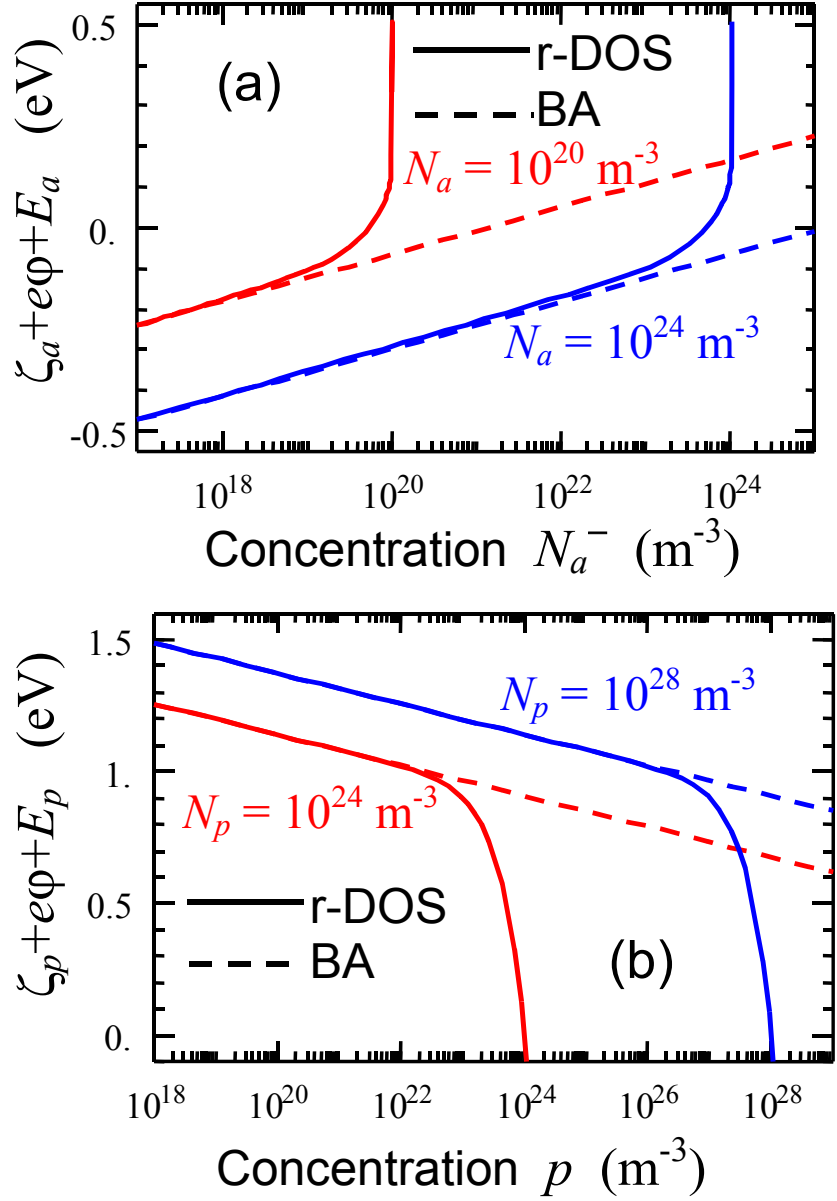

**Fig. 5.** Concentration dependences of ionized acceptors (a) and holes (b) Fermi quasi-levels calculated for (a) $N_a = 10^{20}$ m$^{-3}$ and $10^{24}$ m$^{-3}$; (b) $N_p = 10^{24}$ m$^{-3}$ and $10^{28}$ m$^{-3}$. Other parameters: $T = 293$ K, $\delta E_p = 1$ eV, $\delta E_a = 0.1$ eV. Solid and dashed curves represent the Fermi quasi-levels calculated from expressions (21) for rectangular DOS and Boltzmann approximation for correspondingly.

Under negligibly small impact of electron-hole recombination-generation process (that is typically true at room temperatures without photo-ionization) charges conservation equations are:

$$-\frac{\partial N_a^-}{\partial t} + \frac{1}{e}\frac{\partial J_a}{\partial z} = -\frac{\partial N_a^-}{\partial t} + \frac{\eta_a}{e}\frac{\partial}{\partial z}\left(N_a^- \frac{\partial \zeta_a}{\partial z}\right) = \gamma_a\left(N_a^-, p\right), \qquad (22a)$$

$$\frac{\partial p}{\partial t} + \frac{1}{e}\frac{\partial J_p}{\partial z} = \frac{\partial p}{\partial t} + \frac{\eta_p}{e}\frac{\partial}{\partial z}\left(p\frac{\partial \zeta_p}{\partial z}\right) = -\gamma_a\left(N_a^-, p\right), \tag{22b}$$

$$\frac{\partial N_d^+}{\partial t} + \frac{1}{e}\frac{\partial J_d}{\partial z} = \frac{\partial N_d^+}{\partial t} + \frac{\eta_d}{e}\frac{\partial}{\partial z}\left(N_d^+\frac{\partial \zeta_d}{\partial z}\right) = -\gamma_d\left(N_d^+, n\right), \tag{22c}$$

$$-\frac{\partial n}{\partial t} + \frac{1}{e}\frac{\partial J_n}{\partial z} = -\frac{\partial n}{\partial t} + \frac{\eta_n}{e}\frac{\partial}{\partial z}\left(n\frac{\partial \zeta_n}{\partial z}\right) = \gamma_d\left(N_d^+, n\right). \tag{22d}$$

Here $\gamma_m$ is the carriers hopping function, proportional to the local acceptors-holes and donors-electrons recombination-generation. The simplest model of the functions are $\gamma_a\left(N_a^-, p\right) = \gamma_R^a N_a^- p - \gamma_G^a\left(N_a - N_a^-\right)\theta\left(N_a - N_a^-\right)$, $\gamma_d\left(N_d^+, n\right) = \gamma_R^d N_d^+ n - \gamma_G^d\left(N_d - N_d^+\right)\theta\left(N_d - N_d^+\right)$, where $\theta(x)$ is the step function (compare these expressions with the ones used by Gil et al. [50]). Note, that $N_a \geq N_a^-$ or $N_d \geq N_d^+$ for immobile acceptors or donors respectively, while the sum of Eqs.(22) identically gives the continuity equation $\frac{\partial \rho}{\partial t} + \frac{\partial J_c}{\partial z} = 0$ that does not directly effected by the hopping of carriers.

As it was argued by Riess and Maier [51], equations like (22) should be valid for the local thermal equilibrium and small local gradients, i.e. $J_c^m \sim \eta_m \frac{\partial \zeta_m}{\partial z}$ even when the local situation can be outside the global equilibrium due to the carrier hopping effects. Far from the local equilibrium $J_c^m \sim \eta_m \sinh\left(\frac{\partial \zeta_m}{\partial z}\frac{\delta z}{2k_B T}\right)$ [51]. The hopping terms in the right-hand-side of Eqs.(22) should be obviously taken into account when acceptors (donors) are immobile, but they change their occupation degree (i.e. recharges) dynamically since placed in the time-dependent electric potential, that is in turn created by the time-dependent electric voltage and mobile holes and electrons.

To separate the impact of the carriers hopping and the ions mobility into the charge transfer process, below we analyze two limiting cases:

(1) Assume negligibly small mobility $\eta_{a,d} = 0$ of acceptors (donors), but take into account the hopping process: $\gamma_{a,d} \neq 0$; consequently the hopping currents proportional to $\int^z \gamma_m(z')dz'$ exist in the film.

(2) Assume negligibly small hopping function $\gamma_{a,d} = 0$ of mobile acceptors (donors) with nonzero mobility $\eta_{a,d} \neq 0$.

Material boundary conditions relevant for the considered problem correspond to the limiting cases of the general Chang-Jaffe conditions [15, 16], namely

(a) ultrathin dielectric gap at $z=0$, ions blocking electrode at $z=h$:

$$J_c^a(0) = 0, \quad J_c^a(h) = 0, \quad J_c^d(0) = 0, \quad J_c^d(h) = 0 \qquad (23a)$$

(b) ultrathin dielectric gap at $z=0$, no space charge accumulation at the electrode $z=h$:

$$J_c^p(0) = 0, \quad J_c^n(0) = 0, \quad \rho_S(h) = p(h) - N_a^-(h) - n(h) + N_d^+(h) = 0 \qquad (23b)$$

(c) ultrathin dielectric gaps at $z=0$ and $z=h$:

$$J_c^p(0) = 0, \quad J_c^n(0) = 0, \quad J_c^p(h) = 0, \quad J_c^n(h) = 0, \qquad (23c)$$

(d) alternatively, under the absence of dielectric gap ($H=0$), no space charge accumulation may appear at the conducting electrodes $z=0$ and $z=h$:

$$\rho_S(0) = 0, \quad \rho_S(h) = 0 \qquad (23d)$$

Eqs.(22), (23a), one of Eqs.(23b,c,d) and (14) are the closed form nonlinear boundary problem, which solutions for the case of periodic external voltage change $V(t) = V_0 \sin(\omega t)$ will be analyzed below in linear approximation and nonlinear cases.

Depth distribution of $\varphi$, holes $p(z)/N_p$ and acceptors $N_a^-(z)/N_a$ calculated at four successive moments of cycling (black, red, green and blue curves) are shown in **Figs. 6.** It is seen that the discrepancies between the exact distributions calculated from Eqs.(21)-(22) (solid curves) and Boltzmann approximation (dashed curves) are essential (up to several orders) both at low and high frequency of external voltage cycling. Namely, the concentrations increases without any limitations in the Boltzmann approximation, while the flattening and saturation naturally appear near the film interfaces beyond the Boltzmann approximation. The discrepancies between rectangular DOS and Boltzmann approximation increases with the voltage frequency $w = \omega/2\pi$ increase (compare left and right columns, where the frequency $w\tau_M$ differs in 10 times). Characteristic time is the Maxwell relaxation time is introduced as $\tau_M = \dfrac{R_S^2 e}{\eta_p k_B T} \equiv \dfrac{\varepsilon_{33}^S \varepsilon_0}{\eta_p N_p e}$. The asymmetry of the distributions, seen from the differences between the black and green, red and blue curves generated for symmetrical moments of cycling in **Figs. 6**, originated from the asymmetry of the "prescribed current - prescribed density" boundary conditions $J_c^{a,p}(0) = 0$ and $\rho_S(h) = p(h) - N_a^-(h) = 0$. In the symmetrical "current blocking case", when $J_c^{a,p}(0) = J_c^{a,p}(h) = 0$, the asymmetry disappears (not shown in the figure).

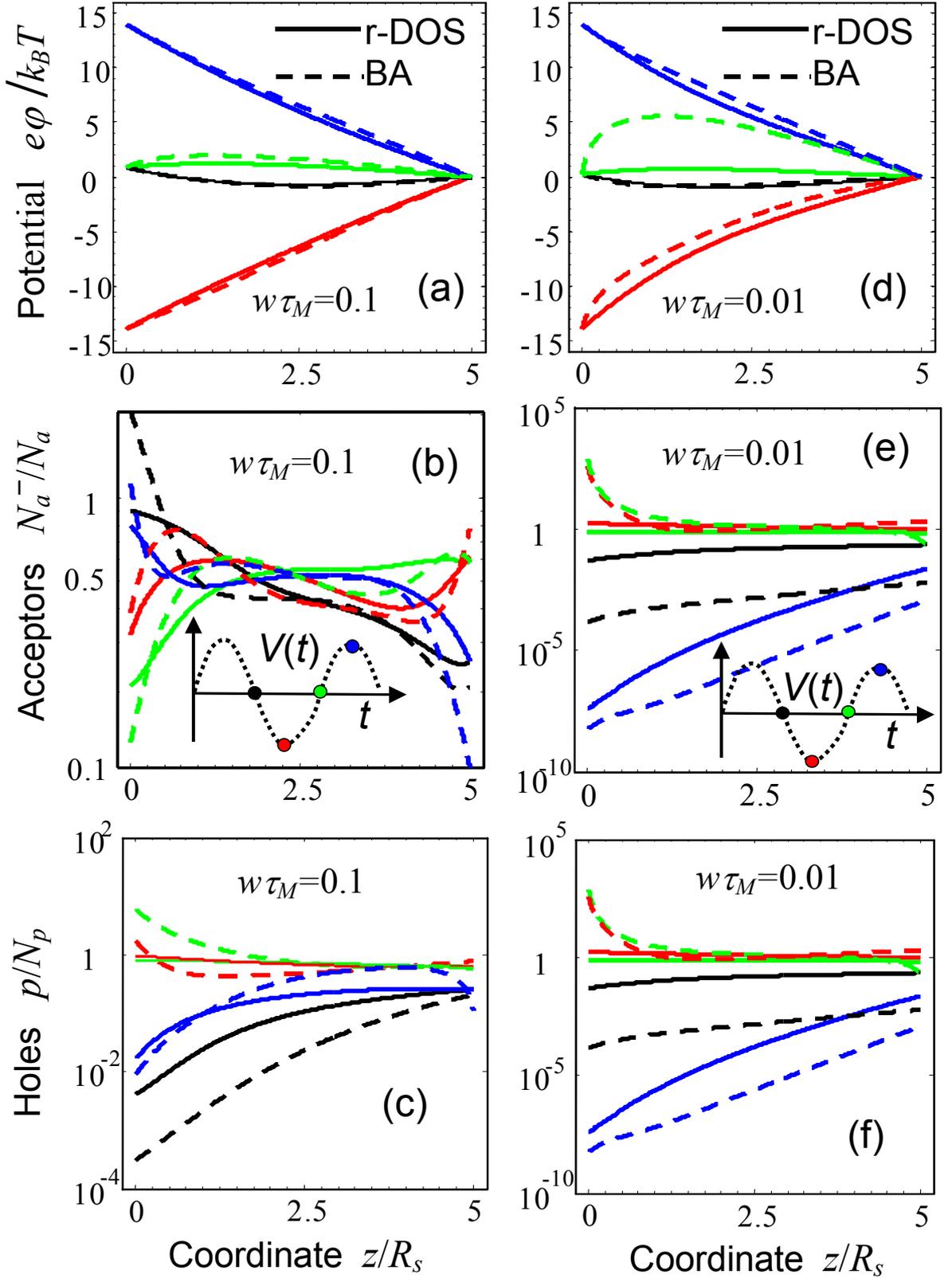

**Figs. 6.** Evolution of the potential φ, holes $p(z)/N_p$ and acceptors $N_a^-(z)/N_a$ distributions calculated at four successive moments of cycling (black, red, green and blue curves) for different frequencies of applied voltage: $w\tau_M = 0.1$ (a, b, c) and $w\tau_M = 0.01$ (d, e, f). DOS parameters $\delta E_a/k_BT = 2$, $\delta E_p/k_BT = 20$, $N_a = N_p$; film thickness $h/R_S = 5$, mobilities ratio $\eta_a/\eta_p = 0.1$

and $\gamma_a(N_a^-, p) = 0$. Solid and dashed curves are generated using rectangular DOS and Boltzmann approximation correspondingly for the chemical potential (21). Insets in (b,e) shows the four successive moments of cycling (black, red, green and blue points). Mixed boundary conditions "prescribed current- prescribed charge density", $J_c^{a,p}(0) = 0$ and $\rho_S(h) = 0$, are imposed.

## 5.2. Space charge and current frequency spectra in the linear drift-diffusion model. Boltzmann approximation for electrons (holes) and immobile ions

One of the important sequences of the 1D-equations $\frac{\partial E_z}{\partial z} = \frac{\rho}{\varepsilon_0 \varepsilon_{33}^S}$ and $\frac{\partial}{\partial z}\left(\varepsilon_0 \varepsilon_{33}^S \frac{\partial E_z}{\partial t} + J_c(t)\right) = 0$, is the possibility to estimate the linear response to the small periodic changes of external voltage $U = U^* + V_0 \exp(i\omega t)$ with frequency $\omega$, until the electrochemical potential derivatives on the carriers concentrations could be regarded constant for the small changes of voltage $V_0$ and finite frequencies. Under these conditions the *linear steady-state* boundary problem for the periodic part of potential $\varphi(z,t) = \varphi_S(z) + \varphi_\omega(z)\exp(i\omega t)$, conductivity current $J_c(z,t) = J_\omega^c(z)\exp(i\omega t)$, the space charge density variations $\rho(z,t) = \rho_S(z) + \rho_\omega(z)\exp(i\omega t)$ and $\sigma_f(t) = \sigma_S + \sigma_\omega \exp(i\omega t)$ acquires the form:

$$\begin{cases} \dfrac{d^2\varphi_\omega}{dz^2} = -\dfrac{\rho_\omega}{\varepsilon_0 \varepsilon_{33}^S}, \quad \dfrac{d}{dz}\left(J_\omega^c - i\omega\varepsilon_0\varepsilon_{33}^S \dfrac{d\varphi_\omega}{dz}\right) = 0, \quad 0 < z < h, \\ \dfrac{d^2\varphi_\omega}{dz^2} = 0, \quad -H < z < 0 \end{cases} \qquad (24)$$

Eqs.(24) should be supplemented with the boundary conditions for the potential and currents at the semiconductor/gap interface:

$$\varphi_\omega(-H) = V_0, \qquad \text{(metal tip-dielectric gap)} \qquad (25a)$$

$$\varphi_\omega(-0) - \varphi_\omega(+0) \approx 0, \quad \text{(semiconductor-dielectric gap)} \qquad (25b)$$

$$D_{2n} - D_{1n} = \sigma_\omega \quad \Rightarrow \quad -\varepsilon_0\left(\varepsilon_{33}^S \frac{d\varphi_\omega(+0)}{dz} - \varepsilon_{33}^g \frac{d\varphi_\omega(-0)}{dz}\right) = \sigma_\omega. \qquad (25c)$$

$$\frac{\partial D_{2n}}{\partial t} + J_{2n}^c = \frac{\partial D_{1n}}{\partial t} + J_{1n}^c \quad \Rightarrow \quad -i\omega\varepsilon_0\varepsilon_{33}^S \frac{d\varphi_\omega(+0)}{dz} + J_\omega^c(0) = -i\omega\varepsilon_0\varepsilon_{33}^g \frac{d\varphi_\omega(-0)}{dz} \qquad (25d)$$

and the condition of the potential vanishing at the earthed electrode $\varphi_\omega(h) = 0$. The compatibility of Eq.(25c) with (25d) leads to the condition $J_\omega^c(0) = i\omega\sigma_\omega = 0$.

Material boundary conditions follows from Eqs.(23): $J_\omega^c(0)=0$, $J_\omega^c(h)=0$ for a dielectric gap at $z=0$ and charge blocking electrode at $z=h$; $J_\omega^c(0)=0$, $\rho_\omega(h)=0$ for a dielectric gap at $z=0$ and charge conducting electrode at $z=h$; $\rho_\omega(0)=0$, $\rho_\omega(h)=0$ for charge conducting electrodes at $z=0$ and $z=h$.

Potential distribution inside the gap $\varphi(z)=V_0+\dfrac{H+z}{\varepsilon_0\varepsilon_{33}^g}\varepsilon_0\varepsilon_{33}^S\dfrac{\partial\varphi(+0)}{\partial z}$ was obtained neglecting the terms proportional to the ratio $\omega^2/c^2$, which is obviously valid in the working frequency range $\omega^2 \ll (c^2/\varepsilon_{33}^g R_S^2) \sim 10^{32}$ 1/s$^2$ without any significant loss of precision.

In the linear drift-diffusion theory the material equation for the current density could be derived in the Boltzmann-Plank-Nernst limit of Eq.(21): $J_\omega^c = -\lambda\dfrac{d}{dz}\varphi_\omega - D\dfrac{d}{dz}\rho_\omega$. At that the static conductivity $\lambda$ and effective diffusion coefficient $D$ are regarded constants: $D=\eta\dfrac{k_B T}{e}$ and $\lambda = \varepsilon_0\varepsilon_{33}^S\dfrac{D}{R_S^2}$, where the Debye screening radius $R_S$ is given by Eq.(18). Then the continuity equation (i.e. the sum of Eqs.(22)) acquires the form $i\omega\rho_\omega + \dfrac{d}{dz}J_\omega^c = 0$.

Note, that the linear material equation $J_\omega^c = -\lambda\dfrac{d}{dz}\varphi_\omega - D\dfrac{d}{dz}\rho_\omega$ is valid for the one prevailing type of carriers, e.g. when *acceptor mobility is absent or much smaller than the holes one,* donors are almost absent and thus the concentration of the free electrons is also negligible in comparison with the holes concentration.

Using the material equation and Eqs.(24) we expressed the space charge density and current via the potential $\varphi_\omega$ as: $J_\omega^c = -\lambda\dfrac{d\varphi_\omega}{dz}+\varepsilon_0\varepsilon_{33}^S D\dfrac{d^3\varphi_\omega}{dz^3}$, $\rho_\omega = -\varepsilon_0\varepsilon_{33}^S\dfrac{d^2\varphi_\omega}{dz^2}$. Then, similarly to Eqs.(8), Eqs.(24)-(25) reduces to the boundary problem inside the semiconductor film:

$$\begin{cases} \dfrac{d^2\varphi_\omega}{dz^2} = -\dfrac{\rho_\omega}{\varepsilon_0\varepsilon_{33}^S} = \dfrac{1}{\varepsilon_0\varepsilon_{33}^S}\dfrac{dJ_\omega^c}{i\omega dz} \rightarrow \left(i\omega + \dfrac{\lambda}{\varepsilon_0\varepsilon_{33}^S}\right)\dfrac{d^2\varphi_\omega}{dz^2} - D\dfrac{d^4\varphi_\omega}{dz^4} = 0, \\ \varphi_\omega(0) - H\dfrac{\varepsilon_{33}^S}{\varepsilon_{33}^g}\dfrac{d\varphi_\omega(0)}{dz} = V_0, \quad \varphi_\omega(h)=0, \\ J_\omega^c(0) = -\lambda\dfrac{d\varphi_\omega(0)}{dz} + \varepsilon_0\varepsilon_{33}^S D\dfrac{d^3\varphi_\omega(0)}{dz^3} = 0, \\ J_\omega^c(h) = -\lambda\dfrac{d\varphi_\omega(h)}{dz} + \varepsilon_0\varepsilon_{33}^S D\dfrac{d^3\varphi_\omega(h)}{dz^3} = 0 \quad \text{or} \quad \rho_\omega(h) = -\varepsilon_0\varepsilon_{33}^S\dfrac{d^2\varphi_\omega(h)}{dz^2} = 0 \end{cases} \quad (26)$$

Characteristic equation corresponding to the linear forth order differential equation (26) has the form $k^2\left(i\omega + \dfrac{\lambda}{\varepsilon_0\varepsilon_{33}^S} - Dk^2\right) = 0$. The roots have the form:

$$k_{1,2}(\omega) = \pm\sqrt{\dfrac{i\omega}{D} + \dfrac{\lambda}{\varepsilon_0\varepsilon_{33}^S D}} = \sqrt{\dfrac{i\omega}{D} + \dfrac{1}{R_S^2}}, \quad k_{3,4}(\omega) = 0. \tag{27}$$

Note, that the first term in $k_{3,4}$ originates from the diffusion contribution, the second one is caused by the linear Debye screening (and thus it is remained in the static case).

The solution of Eq.(25) acquires the form: $\varphi_\omega(z) = \sum_{i=1,2} C_i \exp(k_i z) + C_3 z + C_4$

$\rho_\omega(z) = -\varepsilon_0\varepsilon_{33}^S \sum_{i=1,2} k_i^2 C_i \exp(k_i z)$, and $J_\omega^c(z) = -\lambda C_3 + \sum_{i=1,2}\left(\varepsilon_0\varepsilon_{33}^S D k_i^3 - \lambda k_i\right) C_i \exp(k_i z)$, where $k_{1,2}$ are given by Eq.(27) and four constants $C_i$ should be determined from the boundary conditions to Eq.(26).

After elementary transformations we obtained analytical expressions for all quantities, most important of them are the space charge, conductivity and full current, which are summarized in the **Table 2,** where we introduced the designation for the renormalized gap thickness $\widetilde{H} = \dfrac{\varepsilon_{33}^S}{\varepsilon_{33}^g} H$ and the spatial scale $k(\omega) = \sqrt{\dfrac{i\omega}{D} + \dfrac{1}{R_S^2}}$. Note, that the tunneling current in the gap is regarded negligibly small in comparison with the conductivity current.

**Table 2.** Spectral density of the electric field $E_\omega^z(z)$, conductivity current $J_\omega^c(z)$ and full current $J_\omega^f$, space charge density $\rho_\omega(z)$ and total value $q_\omega = \int_0^h dz\,\rho_\omega(z)$ calculated in the ionic semiconductor film for different boundary conditions.

| | **Boundary conditions** $J_\omega^c(0) = 0, \quad J_\omega^c(h) = 0$ | **Boundary conditions** $J_\omega^c(0) = 0, \quad \rho_\omega(h) = 0$ |
|---|---|---|
| $E_\omega^z$ | $\dfrac{V_0}{h+\widetilde{H}} \cdot \dfrac{i\omega\varepsilon_0\varepsilon_{33}^S + \lambda\dfrac{e^{k(h-z)} + e^{kz}}{1+e^{2kh}}}{i\omega\varepsilon_0\varepsilon_{33}^S + \lambda\dfrac{\tanh(hk/2) + k\widetilde{H}}{(h+\widetilde{H})k}}$ | $\dfrac{V_0}{h+\widetilde{H}} \cdot \dfrac{i\omega\varepsilon_0\varepsilon_{33}^S + \lambda\dfrac{e^{k(2h-z)} + e^{kz}}{1+e^{2kh}}}{i\omega\varepsilon_0\varepsilon_{33}^S + \lambda\dfrac{\tanh(hk) + k\widetilde{H}}{(h+\widetilde{H})k}}$ |
| $J_\omega^c$ | $\dfrac{V_0}{h+\widetilde{H}} \dfrac{i\omega\varepsilon_0\varepsilon_{33}^S\lambda\left(1+e^{kh} - e^{k(h-z)} - e^{kz}\right)}{\left(1+e^{kh}\right)\left(i\omega\varepsilon_0\varepsilon_{33}^S + \lambda\dfrac{2\tanh(hk/2) + k\widetilde{H}}{(h+\widetilde{H})k}\right)}$ | $\dfrac{V_0}{h+\widetilde{H}} \cdot \dfrac{i\omega\varepsilon_0\varepsilon_{33}^S\lambda\left(1+e^{2kh} - e^{k(2h-z)} - e^{kz}\right)}{\left(1+e^{2kh}\right)\left(i\omega\varepsilon_0\varepsilon_{33}^S + \lambda\dfrac{\tanh(hk) + k\widetilde{H}}{(h+\widetilde{H})k}\right)}$ |
| $J_\omega^f$ | $\dfrac{V_0}{h+\widetilde{H}} \cdot \dfrac{i\omega\varepsilon_0\varepsilon_{33}^S (i\omega\varepsilon_0\varepsilon_{33}^S + \lambda)}{i\omega\varepsilon_0\varepsilon_{33}^S + \lambda \dfrac{2\tanh(hk/2)+k\widetilde{H}}{(h+\widetilde{H})k}}$ | $\dfrac{V_0}{h+\widetilde{H}} \cdot \dfrac{i\omega\varepsilon_0\varepsilon_{33}^S (i\omega\varepsilon_0\varepsilon_{33}^S + \lambda)}{i\omega\varepsilon_0\varepsilon_{33}^S + \lambda \dfrac{\tanh(hk)+k\widetilde{H}}{(h+\widetilde{H})k}}$ |
| $\rho_\omega$ | $\dfrac{V_0}{h+\widetilde{H}} \cdot \dfrac{\varepsilon_0\varepsilon_{33}^S \lambda \cdot k\left(e^{kz}-e^{k(h-z)}\right)}{(1+e^{kh})\left(i\omega\varepsilon_0\varepsilon_{33}^S + \lambda \dfrac{2\tanh(hk/2)+k\widetilde{H}}{(h+\widetilde{H})k}\right)}$ | $\dfrac{V_0}{h+\widetilde{H}} \cdot \dfrac{\varepsilon_0\varepsilon_{33}^S \lambda \cdot k\left(e^{kz}-e^{k(2h-z)}\right)}{(1+e^{2kh})\left(i\omega\varepsilon_0\varepsilon_{33}^S + \lambda \dfrac{\tanh(hk)+k\widetilde{H}}{(h+\widetilde{H})k}\right)}$ |
| $q_\omega$ | Zero in linear approximation. No total strain-voltage response could be detected in accordance with Eq.(30). | $\dfrac{V_0}{h+\widetilde{H}} \cdot \dfrac{\varepsilon_0\varepsilon_{33}^S \lambda (1-e^{kh})^2}{(1+e^{2kh})\left(i\omega\varepsilon_0\varepsilon_{33}^S + \lambda \dfrac{\tanh(hk)+k\widetilde{H}}{(h+\widetilde{H})k}\right)}$ |
| P.S. | Note, that for the case of boundary conditions, $\rho_\omega(0) = \rho_\omega(h) = 0$, we derived that $E_\omega^z = \dfrac{V_0}{h+\widetilde{H}}$, $J_\omega^c = \dfrac{\lambda V_0}{h+\widetilde{H}}$, $J_\omega^f = \dfrac{V_0}{h+\widetilde{H}}(i\omega\varepsilon_0\varepsilon_{33}^S + \lambda)$, at that both local and total electroneutrality holds: $\rho_\omega = q_\omega = 0$. Thus no total strain-voltage response could be detected in the linear approximation according to Eq.(30). | |

It is seen from the **Table 2** that the linear current spectra, calculated for different boundary conditions, are the same for two limiting cases, namely at high frequencies (for any thickness) and at high thickness (for not very low frequencies). At the same time, at low frequencies the current spectra are either capacitor like ($J_\omega^c(0) = J_\omega^c(h) = 0$ and $J_\omega^c(0) = \rho_\omega(h) = 0$ conditions) or resistor-like ($\rho_\omega(0) = \rho_\omega(h) = 0$ conditions). Also note the qualitative similarity between current spectra for $J_\omega^c(0) = J_\omega^c(h) = 0$ and $J_\omega^c(0) = \rho_\omega(h) = 0$ conditions at arbitrary frequencies. In the next subsections we will show that these results remain qualitatively valid for nonlinear solutions.

Z-distributions of the electric potential $\varphi_\omega(z)$, field $E_\omega^z(z)$, space charge $\rho_\omega(z)$ and conductivity current $J_\omega^c(z)$ are shown in **Figs. 7** for the mixed-type boundary conditions $J_\omega^c(0) = \rho_\omega(h) = 0$ and fixed frequency *w*.

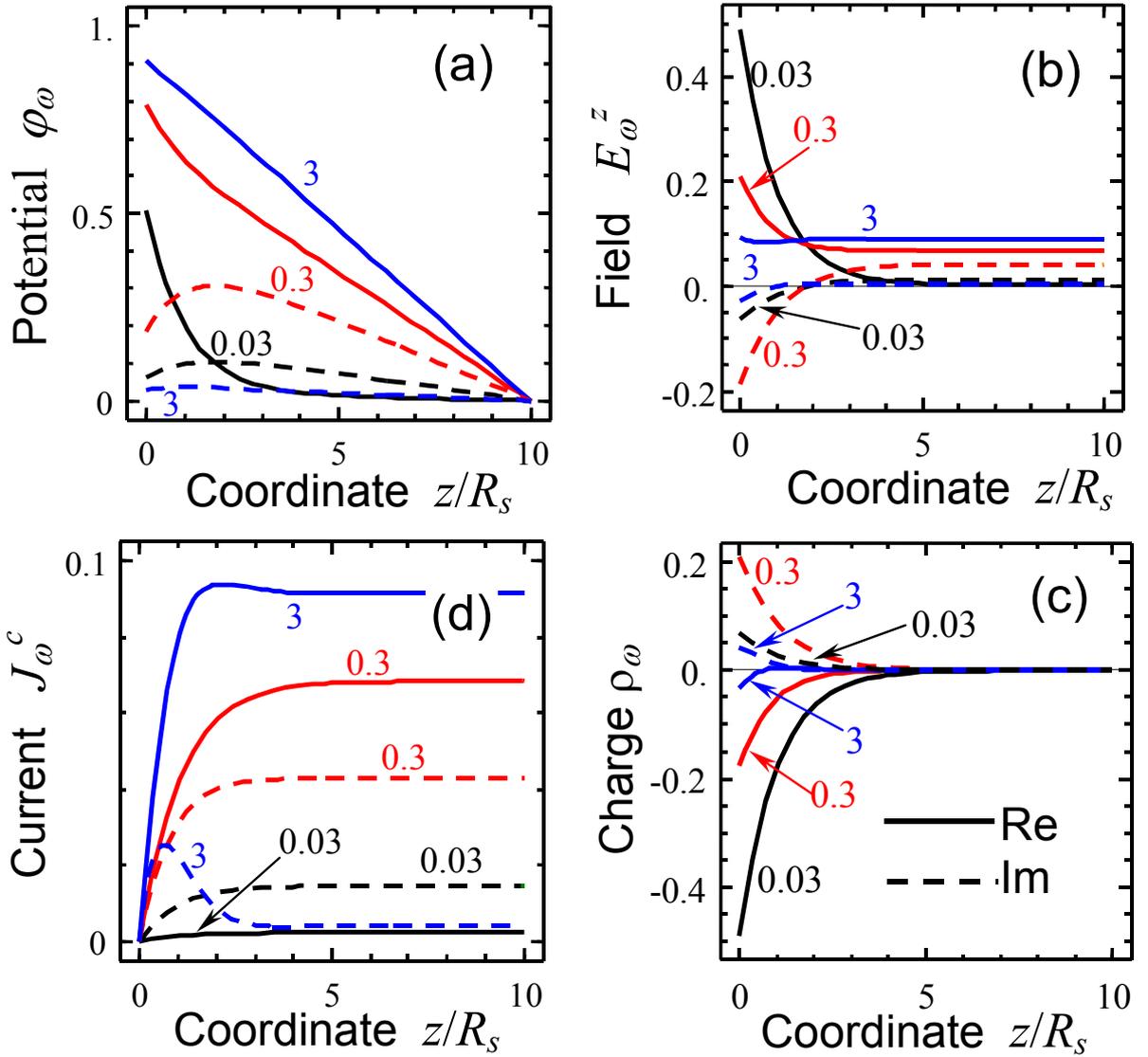

**Fig. 7.** Distributions of the electric potential $\frac{\varphi_\omega(z)}{V_0}$ (a), field $\frac{E_\omega^z(z)R_S}{V_0}$ (b), space charge $\frac{\rho_\omega(z)R_S}{V_0\varepsilon_0\varepsilon_{33}^S}$ (c) and conductivity current $\frac{J_\omega^c(z)R_S}{V_0\lambda}$ (d) calculated for dimensionless frequency values $w\tau_M$ =0.03, 0.3, 3 (figures near the curves), gap thickness $\widetilde{H}/R_S = 1$, film thickness $h/R_S = 10$, boundary conditions $J_\omega^c(0) = \rho_\omega(h) = 0$. Solid and dashed curves represent real and imaginary parts.

Despite the conductivity current spectra $J_\omega^c$ is z-dependent, the full current spectra $J_\omega^f$ is independent on the coordinate z as anticipated from the conservation law. Dependences of the full current $J_\omega^f$ and total space charge $q_\omega$ vs. dimensionless frequency $w\tau_M$ are shown in **Figs. 8** for several gap thicknesses and mixed-type boundary conditions $J_\omega^c(0) = \rho_\omega(h) = 0$ (here $\tau_M = \varepsilon_0 \varepsilon_{33}^S / \lambda \equiv R_S^2 / D$ is the relaxation time). In the limiting case of zero gap ($H \to 0$) the current and charge are maximal; they decrease with the gap thickness increase. The current absolute value increases with frequency increase, has a plateau and becomes proportional to ω at high frequencies [**Figs. 8a,b**]. The total charge absolute value monotonically decreases with frequency increase; while its imaginary part has maximum as shown in **Figs. 8c,d**.

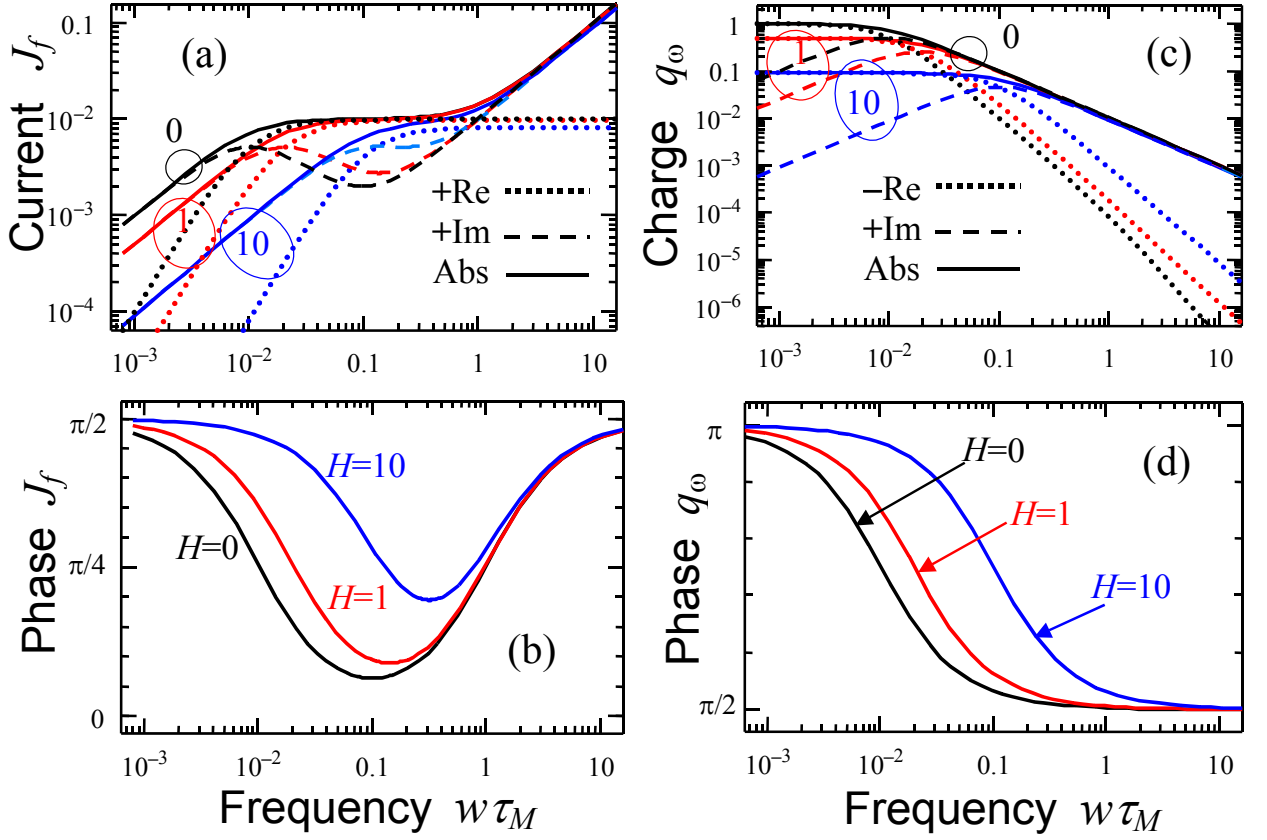

**Fig. 8.** Real, imaginary parts, absolute value (dotted, dashed and solid curves) and phase (b,d) of the full current $J_\omega^f \dfrac{R_S}{V_0 \lambda}$ (a,b) and total space charge $\dfrac{q_\omega R_S^2}{V_0 \varepsilon_0 \varepsilon_{33}^S}$ (c,d) vs. dimensionless frequency $w$ calculated for several gap thickness $\widetilde{H}/R_S = 0, 1, 10$ (figures near the curves). Film thickness $h/R_S = 100$, boundary conditions $J_\omega^c(0) = \rho_\omega(h) = 0$.

Note, that the conventional local charge density approximation $\rho_\omega(z) = -\varepsilon_0 \varepsilon_{33}^S \dfrac{\varphi_\omega(z)}{R_S^2}$ leads to the independence of the potential distribution on the external field frequency (see **Appendix C**). The current spectra $J_f(z, \omega)$ appeared simply proportional to the product $i\omega$, which corresponds to the pure capacitive reactance. The result looks unrealistic for existing semiconductors especially at high frequencies. So, at least for the high-frequency case, the non-locality of the current response to external periodic voltage should be considered. Actually the expressions for $\rho_\omega(z)$, listed in the **Table 2**, are not proportional to the potential distribution, but contain the non-local contribution originated from the complex dependence in the spatial scale $k(\omega) = \sqrt{\dfrac{i\omega}{D} + \dfrac{1}{R_S^2}}$.

*5.3. Nonlinear current – voltage characteristics at finite frequencies: calculations in the Boltzmann approximation and for rectangular DOS*

Below we analyze the dynamic current-voltage response caused by the *mobile* ionized acceptors and holes in the ionic-semiconductor film and *negligibly small impact of the local generation-recombination of holes*. Note, that the dynamic current-voltage response caused by the mobile ionized donors and electrons and corresponding currents can be analyzed in a similar way.

Typical current – voltage characteristics, $J_f(V)$, were calculated numerically in the frequency range $w\tau_M = 0.001–0.1$ of external voltage $V(t) = V_0 \sin(\omega t)$ and several voltage amplitude $V_0$ (see different loops in **Figs. 9-11**). Hereinafter we re-introduce the Maxwell relaxation time $\tau_M = \dfrac{\varepsilon_0 \varepsilon_{33}^S}{\eta_p N_p e}$, linear frequency $w = \omega/2\pi$ and dimensionless currents $\widetilde{J}_a^c = \dfrac{\tau_M J_a^c}{R_S N_p e}$, $\widetilde{J}_p^c = \dfrac{\tau_M J_p^c}{R_S N_p e}$, where $\dfrac{1}{R_S} = \sqrt{\dfrac{e^2 N_p}{\varepsilon_{33}^S \varepsilon_0 k_B T}}$. As anticipated, we obtained that the full current is spatially homogeneous with accuracy of not more than 1% numerical error (not shown in the figures). Dimensionless system of equations (14) and (22) is given in **Appendix B**.

The **Figs. 9-11,** calculated without hopping contribution ($\gamma_a(N_a^-, p) = 0$), demonstrate principally different loop shapes, at that the differences mainly originate from the type of boundary conditions. Namely:

(I) The current – voltage loops, shown in **Figs. 9**, are calculated for the case of symmetric "holes conducting" ($\rho_S(0) = \rho_S(h) = 0$) and "ion blocking" ($J_c^a(0) = J_c^a(h) = 0$) boundary conditions. They are symmetric with respect to the voltage sign. At low frequencies $w\tau_M \leq 0.001$ the loops transforms into the nonlinear I-V characteristic. At frequencies $w\tau_M \sim 0.01$ their shape is ellipsoidal at small voltage $V_0$, with $V_0$ increase they mimic slim hysteresis loops for rectangular DOS or demonstrate "resistive switching" double loops in the Boltzmann approximation. Then the loops becomes noticeably inflated with the frequency increase $w\tau_M \geq 0.1$, but the inflation is stronger in the Boltzmann approximation.

(II) The current – voltage loops, shown in **Figs. 10**, are calculated for the case of symmetric "carriers blocking" boundary conditions: $J_c^{a,p}(0) = J_c^{a,p}(h) = 0$. They are symmetric with respect to the voltage sign. For the case the average density of holes and acceptors are time independent, despite the distributions changes in time. The loops are much overblown in comparison with the case (I). The current value decreases with the frequency decrease at tends to zero in the static limit as anticipated for nonzero gap acting as plain capacitor. At low frequencies $w\tau_M \leq 0.001$ loops shape is quasi-circular at small voltages $V_0$, with $V_0$ increase they becomes parallelogram-like for rectangular DOS and four petal-shaped in the Boltzmann approximation. The loop shape becomes circular with the frequency increase $w\tau_M \geq 0.1$.

(III) The current – voltage loops, shown in **Figs. 11**, are calculated for the case of asymmetric mixed-type boundary conditions: $J_c^a(0) = J_c^p(0) = 0$ and $\rho_S(h) = 0$, $J_c^a(h) = 0$. They are strongly asymmetric with respect to the voltage sign and have irregular shape for low frequencies $w\tau_M \leq 0.01$, both calculated for rectangular DOS and in the Boltzmann approximation. The loops becomes noticeably inflated with the frequency increase $w\tau_M \geq 0.1$, and the inflation is much stronger in the Boltzmann approximation. The current value decreases with the frequency decrease at tends to zero in the static limit.

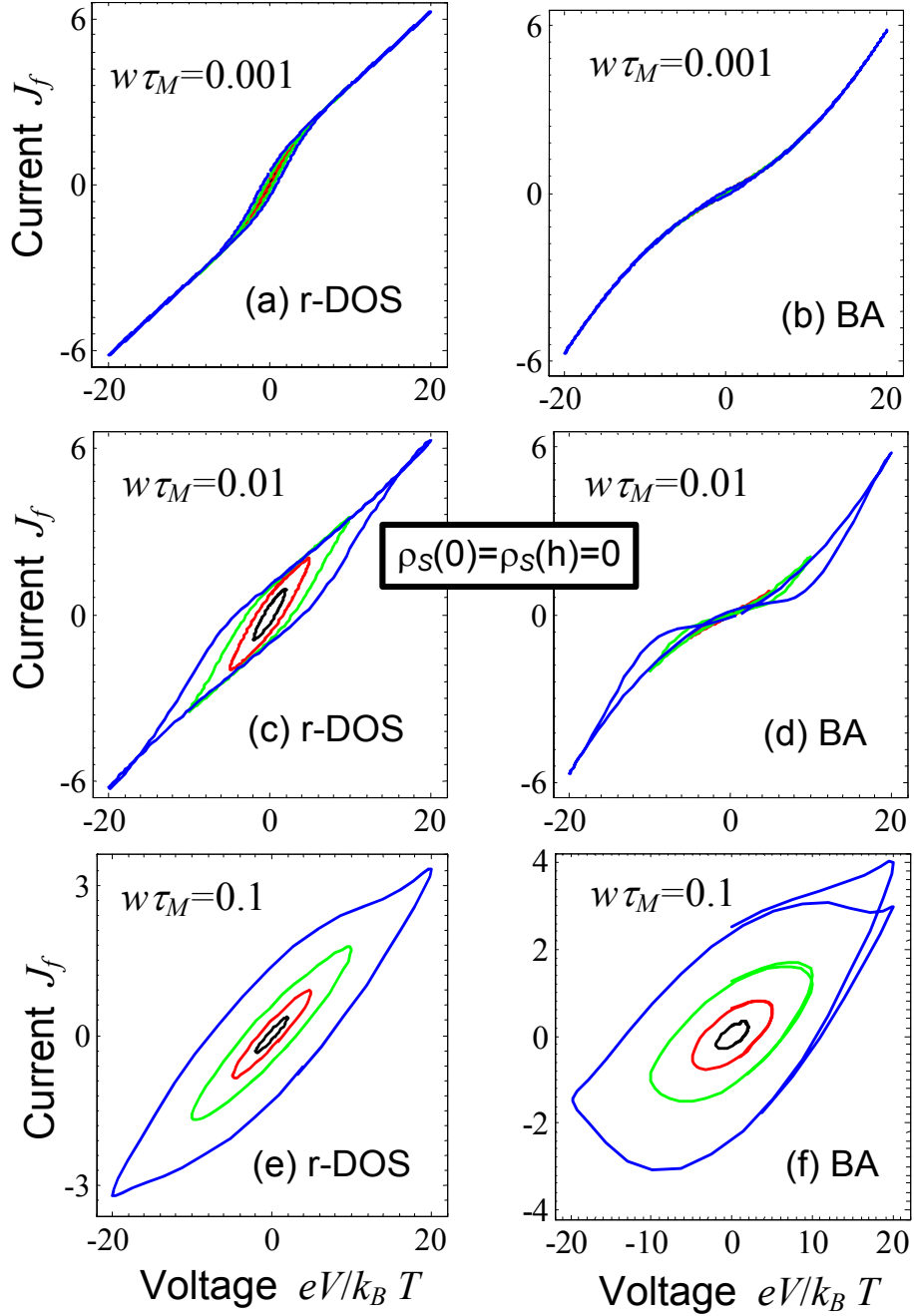

**Fig. 9.** Current – voltage response calculated for different frequencies of external voltage: $w\tau_M = 0.001$ (a, b), $w\tau_M = 0.01$ (c, d) and $w\tau_M = 0.1$ (e, f). Different loops (black, red, green and blue) correspond to the different maximal voltage $V_0 = 2, 5, 10, 20$ (in $k_B T/e$ units). Plots (a, c, e) are generated using rectangular DOS for the chemical potential (21) and plots (b, d, f) are generated in the Boltzmann approximation. DOS parameters $\delta E_a/k_B T = 2$, $\delta E_p/k_B T = 20$, $N_a = N_p$; film thickness $h/R_S = 5$, mobilities ratio $\eta_a/\eta_p = 0.1$ and $\gamma_a(N_a^-, p) = 0$. Symmetric "holes conducting" $\rho_S(0) = \rho_S(h) = 0$ and "ion blocking" $J_c^a(0) = J_c^a(h) = 0$ boundary conditions are imposed. Note, that for the case the gap should be absent ($H = 0$).

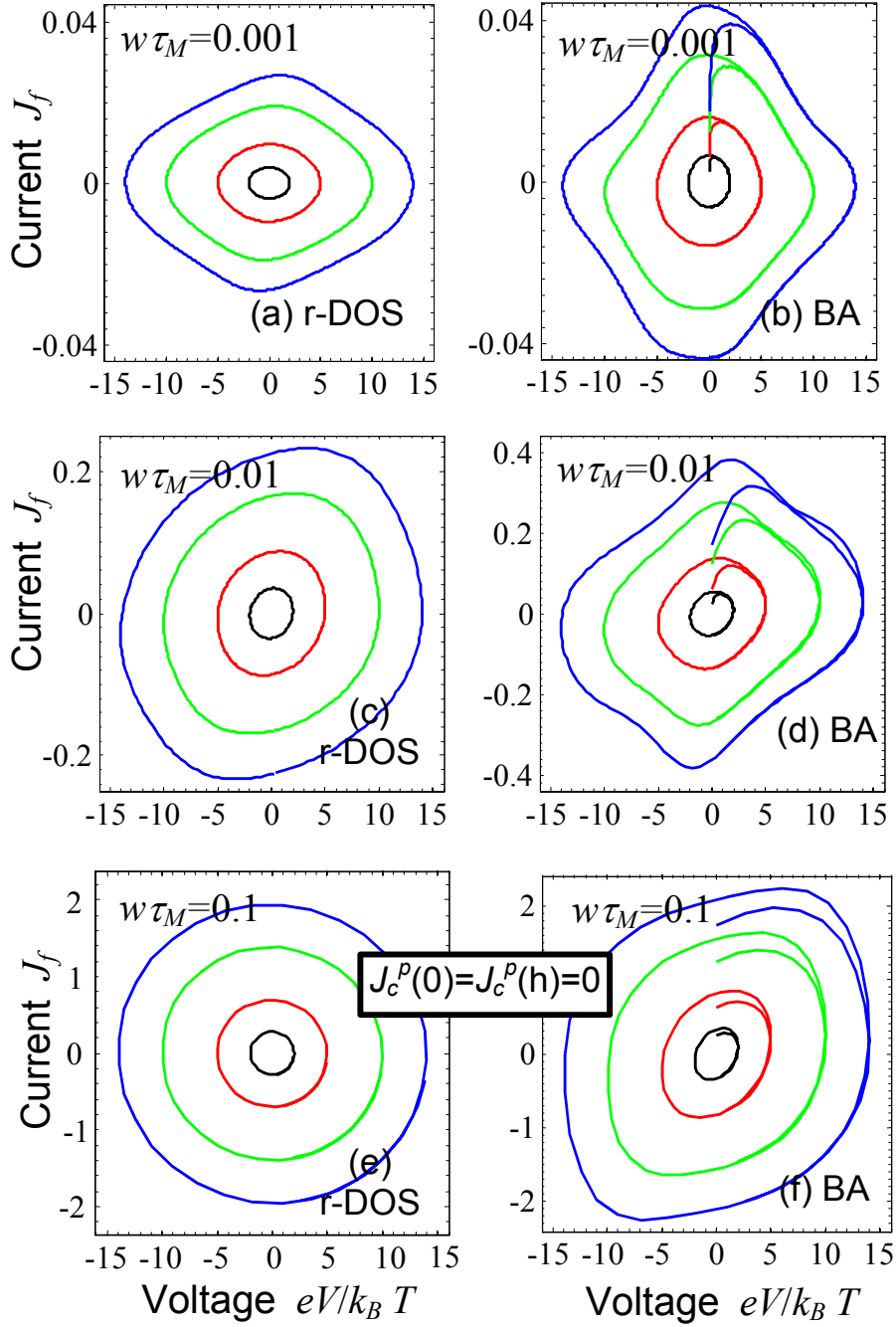

**Fig. 10.** Current − voltage response calculated for different frequencies of external voltage: $w\tau_M = 0.001$ (a, b), $w\tau_M = 0.01$ (c, d) and $w\tau_M = 0.1$ (e, f). Different loops (black, red, green and blue) correspond to the increasing maximal voltage $V_0 = 2, 5, 10, 14$ (in $k_B T/e$ units). Plots (a, c, e) are generated using rectangular DOS for the chemical potential (21); plots (b, d, f) are generated in the Boltzmann approximation. DOS parameters $\delta E_a/k_B T = 1$, $\delta E_p/k_B T = 10$, $N_a = N_p$; film thickness $h/R_S = 5$, mobilities ratio $\eta_a/\eta_p = 0.1$ and $\gamma_a(N_a^-, p) = 0$. Symmetric "carriers blocking" boundary conditions $J_{a,p}^c(0) = J_{a,p}^c(h) = 0$ are imposed.

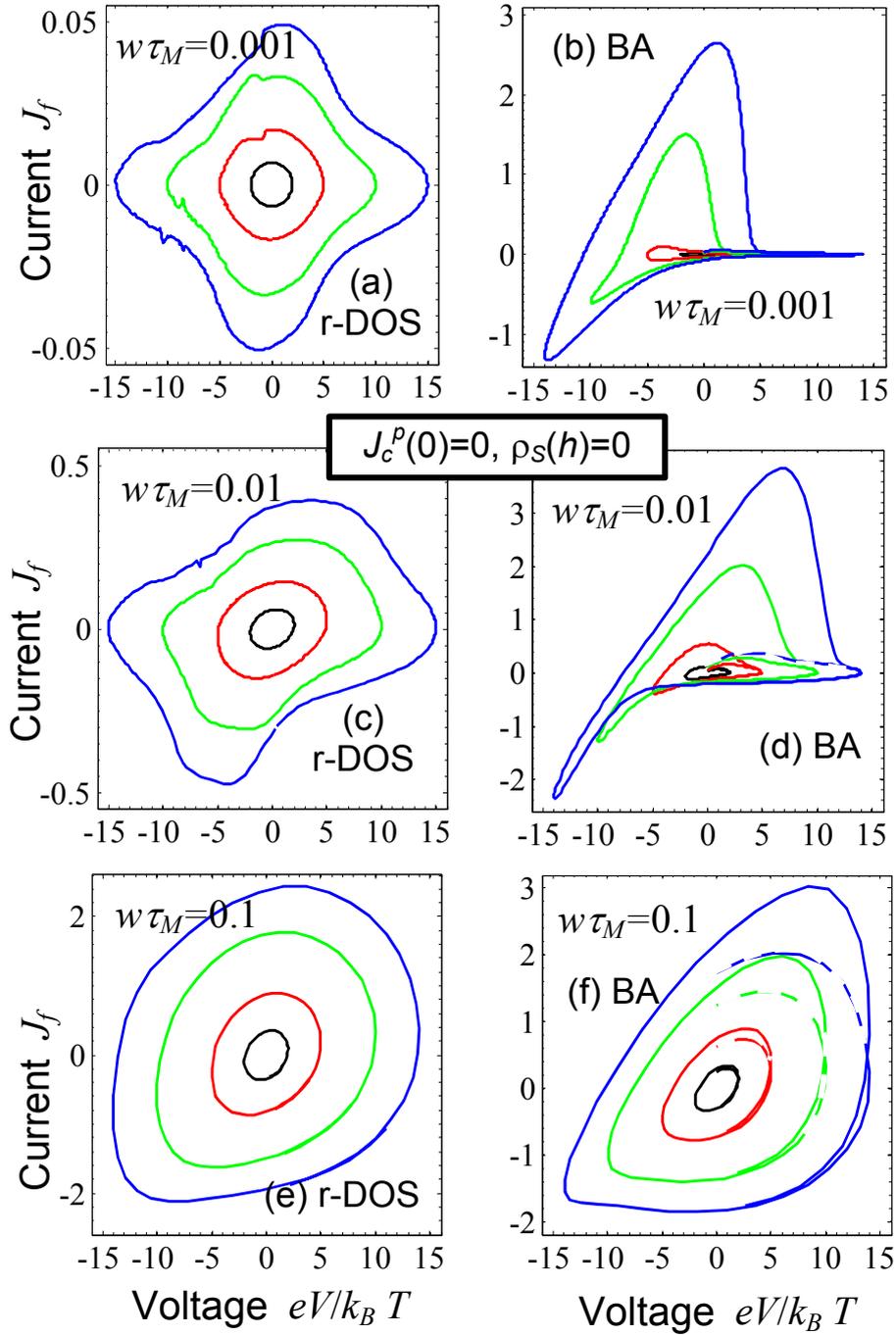

**Fig. 11.** Current − voltage response calculated for different frequencies of external voltage: $w\tau_M = 0.001$ (a, b), $w\tau_M = 0.01$ (c, d) and $w\tau_M = 0.1$ (e, f). Different loops (black, red, green and blue) correspond to the different maximal voltage $V_0 = 2, 5, 10, 14$ (in $k_B T/e$ units). Plots (a, c, e) are generated using rectangular DOS for the chemical potential (21) and plots (b, d, f) are generated in the Boltzmann approximation. DOS parameters $\delta E_a / k_B T = 1$, $\delta E_p / k_B T = 10$, $N_a = N_p$; film thickness $h/R_S = 5$, mobilities ratio $\eta_a / \eta_p = 0.1$ and $\gamma_a(N_a^-, p) = 0$. Asymmetric boundary conditions $\rho_S(h) = 0$, $J_c^p(0) = 0$ and $J_c^a(0) = J_c^a(h) = 0$ are imposed.

Thus, only in the symmetrical cases the asymmetry $V \to -V$ disappears. Also it is seen that the discrepancies between current-voltage loops calculated from Eqs.(21) for Fermi quasi-levels using rectangular DOS (plots a,c,e) and in the Boltzmann approximation (plots b,d, f) are essential regarding the loop shape and sometimes the amplitude, which reaches the several orders of magnitude for the **Figs.11a,b**. The discrepancy between the loop shape and amplitude strongly increases with the voltage amplitude $V_0$ increase (compare left and right columns, top and bottom plots, where the frequency differs in 100 times). Actually, it is seen that the Boltzmann approximation becomes invalid with $V_0$ increase. The explanation follows from the fact that the holes and acceptor concentrations could increase without any limitations in the Boltzmann approximation, while the flattening and saturation naturally appear near the film interfaces for realistic DOS, but it cannot be taken into account in the Boltzmann approximation (see **Figs. 6**).

Finally, let us analyze the dynamic current-voltage response caused by the *immobile* acceptors and mobile holes allowing for *the local generation-recombination of holes* in the ionic-semiconductor film. Typical voltage dependence of the full current, average concentration of holes $\langle p \rangle = \frac{1}{h}\int_0^h dz\, p(z)$ and immobile ionized acceptors $\langle N_a^- \rangle = \frac{1}{h}\int_0^h dz\, N_a^-(z)$ are shown in **Figs. 12** for $\gamma_a(N_a^-, p) \neq 0$ and different types of boundary conditions.

Current-voltage characteristics, shown in **Figs. 12a-c** are ellipse-like for the "holes-conducting" boundary conditions $\rho_S(0) = \rho_S(h) = 0$ and circle-like for the "holes-blocking" conditions $J_c^p(0) = J_c^p(h) = 0$ and "mixed" conditions $\rho_S(h) = 0$, $J_c^p(0) = 0$, since the condition $J_c^p(0) = 0$, originated from the dielectric gap, acts as capacitor at relatively low frequencies. Note, that the dependences shown in **Figs. 12a-c** are rather different from those shown in **Figs. 9-11**, which correspond to the mobile acceptors and $\gamma_a(N_a^-, p) = 0$. Thus, we predict that the principal differences between the cases of mobile and immobile acceptors, as well as hopping conductivity impact, could be distinguished experimentally, at that the differences increases with the external voltage frequency decrease.

For the symmetric boundary conditions $\rho_S(0) = \rho_S(h) = 0$ and $J_c^p(0) = J_c^p(h) = 0$ the average concentrations of holes and ionized acceptors very quickly tend to the voltage independent stationary point of the system (22): $N_a^- = p = \dfrac{2N_a}{1 + \sqrt{1 + 4\gamma_R^a N_a/\gamma_G^a}}$. Moreover the local electroneutrality $\rho_S(z) = 0$ is perfectly holds for the case $\rho_S(0) = \rho_S(h) = 0$ independently on the voltage frequency and amplitude (compare **Figs.12d** with conclusions of the **Table 2**). For the asymmetric mixed-type boundary conditions $\rho_S(h)=0$, $J_c^p(0)=0$ the average concentrations of ionized acceptors and holes are essentially different: the holes concentration appeared strongly dependent on the voltage amplitude and frequency (compare **Figs.12e** with the **Table 2**), the concentration of immobile ionized acceptors is almost independent on applied voltage and its frequency, its value is slightly deviates from the stationary concentration $N_a^- = \dfrac{2N_a}{1+\sqrt{1+4\gamma_R^a N_a/\gamma_G^a}}$. Note, that the local electroneutrality condition is far not true for the mixed-type boundary conditions.

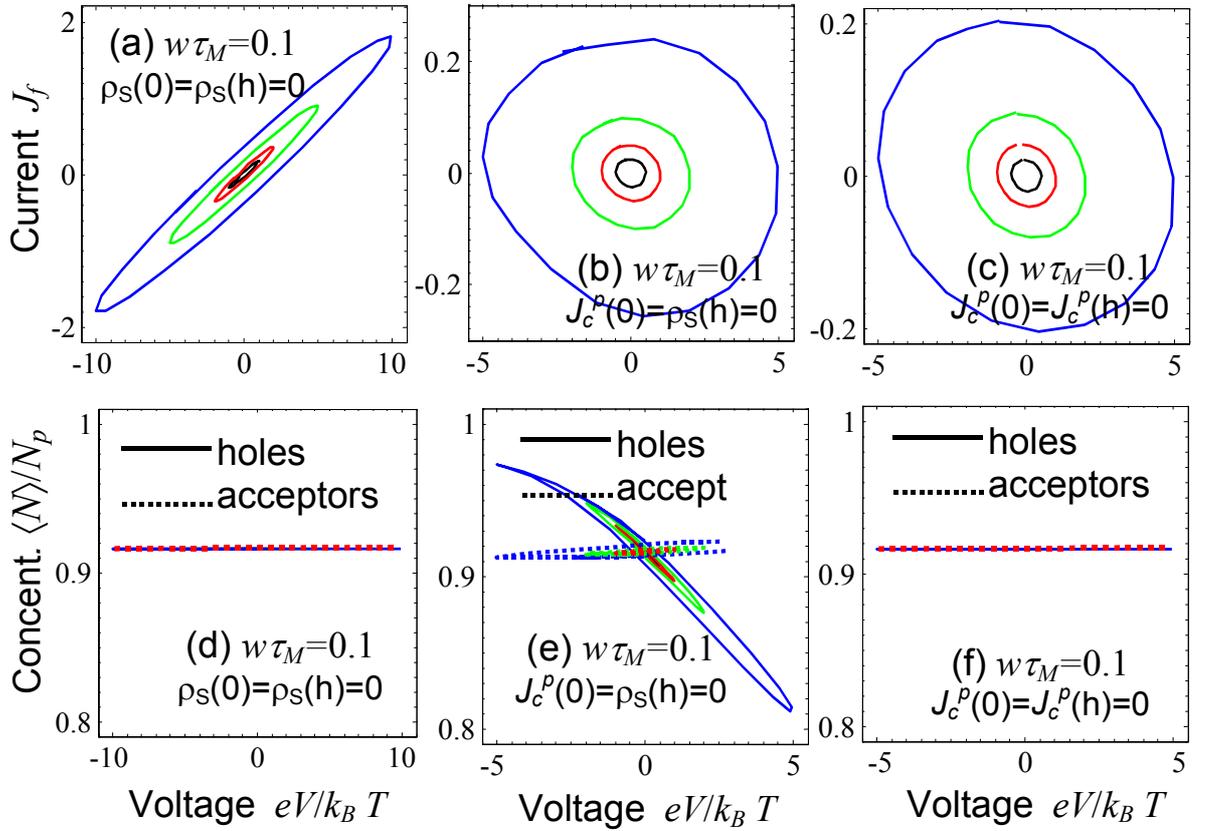

**Fig. 12.** Voltage dependence of the full current (a,b,c) and (d,e,f) average concentration of holes (solid curves) and ionized acceptors (dotted curves) calculated for the frequency $w\tau_M = 0.1$, acceptors mobility $\eta_a = 0$, $\tau_M \gamma_G^a = 0.3$, $\gamma_R^a N_a/\gamma_G^a = 0.1$ and different types of boundary conditions listed in the plots labels. Different loops (black, red, green and blue) correspond to the different maximal voltage in $k_B T/e$ units. Rectangular DOS parameters $\delta E_a/k_B T = 1$, $\delta E_p/k_B T = 10$, $N_a = N_p$; film thickness $h/R_S = 5$.

To summarize the results of the Section 5, we evolve analytical formalism capable to describe the current frequency spectra (linear approximation). Also we predict the great variety of the nonlinear dynamic current-voltage loops of the ionic semiconductor film with mobile acceptors and holes, placed between the SPM tip electrode and substrate electrode with different conductive properties. Some of the calculated loops mimic the characteristics, experimentally observed in the correlated oxides and resistive switching materials [52, 53]. We demonstrated that the conventional Boltzmann approximation becomes invalid with the increase of maximal voltage amplitude $V_0 > k_B T/e$, when the realistic DOS should be used for the correct calculations of the Fermi-quasi levels and generalized fluxes. Note, that the current-voltage response of the ionic semiconductor film with mobile ionized donors and electrons can be analyzed in a similar way. We also predict that the principal differences between the cases of mobile and immobile acceptors, as well as hopping conductivity impact, could be distinguished experimentally by current SPM, at that the differences increases with the external voltage frequency decrease.

## 6. SPM strain-voltage response: calculations in decoupling approximation

As described in the end of the Section 2, the elastic strain of ionic semiconductor films is caused by electric field produced by the SPM tip, which in turn changes the acceptors (donors) occupation degree and causes electromigration of ions, electrons and/or holes in the film. Actually, the equations of state for the elastic media, subjected to the carrier concentration variations $\delta c_m(\mathbf{r})$, mechanical stress tensor $\sigma_{ij}$ and elastic strain $u_{ij}$ are: $u_{ij}(\mathbf{r}) = \sum_m \beta_{ij}^m \delta c_m(\mathbf{r}) + s_{ijkl} \sigma_{kl}(\mathbf{r})$, where $s_{ijkl}$ is the tensor of elastic compliances and $\beta_{ij}^m$ consists of the *ions stochimetric and recharging* contributions ($\beta_{ij}^{a,d}$), and *electron-phonon deformation potential* ($\beta_{ij}^{p,n}$) as was argued in the Section 2 and estimated in the **Table 1**.

The squashed tip electrode is regarded mechanically free (the dielectric gap is thin and mechanically flexible), so the normal stress is absent at $z = 0$. The bottom electrode is clamped or rigid, so here the displacement $u_i$ is fixed at $z = h$. Thus the elastic boundary conditions are

$$\sigma_{3i}(0,t) = 0, \quad u_i(h,t) = 0. \tag{28}$$

Allowing for the boundary conditions (28) and equation $u_{ij}(\mathbf{r}) = \sum_m \beta_{ij}^m \delta c_m(\mathbf{r}) + s_{ijkl} \sigma_{kl}(\mathbf{r})$, the equation of mechanical equilibrium $\partial \sigma_{ij}(z,t)/\partial x_j = 0$ and compatibility conditions have the following solution of the considered 1D-problem [8]:

$$\sigma_{13} = \sigma_{23} = \sigma_{33} = 0, \qquad \sigma_{12} = 0,$$

$$\sigma_{11}(z,t) = \sum_m \left( \frac{s_{12}\beta_{22}^m - s_{11}\beta_{11}^m}{s_{11}^2 - s_{12}^2} \right) \delta c_m(z,t) \approx -\sum_m \frac{\beta_{11}^m \delta c_m(z,t)}{s_{11} + s_{12}}, \qquad (29a)$$

$$\sigma_{22}(z,t) = \sum_m \left( \frac{s_{12}\beta_{11}^m - s_{11}\beta_{22}^m}{s_{11}^2 - s_{12}^2} \right) \delta c_m(z,t) \approx -\sum_m \frac{\beta_{11}^m \delta c_m(z,t)}{s_{11} + s_{12}}.$$

$$u_{11} = u_{12} = u_{13} = u_{23} = u_{22} = 0,$$

$$u_{33}(z,t) = \sum_m \left( \beta_{33}^m - \frac{s_{12}(\beta_{11}^m + \beta_{22}^m)}{s_{11} + s_{12}} \right) \delta c_m(z,t) \approx \sum_m \left( \beta_{33}^m - \frac{2s_{12}\beta_{11}^m}{s_{11} + s_{12}} \right) \delta c_m(z,t). \qquad (29b)$$

The approximate equalities correspond to the almost transversally isotropic tensor $\beta_{11}^m \approx \beta_{22}^m \neq \beta_{33}^m$, the assumption is mainly used hereinafter.

The displacement of the ionic semiconductor film surface, measured by the strain SPM, is calculated from Eq.(29b) as:

$$u_3(0,\omega) = -\sum_m \left( \beta_{33}^m - \frac{s_{12}(\beta_{11}^m + \beta_{22}^m)}{s_{11} + s_{12}} \right) \int_0^h dz\, \delta c_m(z,\omega). \qquad (30)$$

The surface strain-voltage response (30) is caused by the local recharging of acceptors and donors as well as by the electromigration of the free charge carriers coupled with the strain via the tensorial deformation potential $\beta_{ij}^{p,n}$.

### 6.1. Static limit of the nonlinear strain-voltage response

Below we analyze the static strain-voltage response caused by the ionized acceptors and holes in the semiconductor film. Note, that the strain-voltage response caused by the ionized donors and electrons can be analyzed in a similar way.

Concentration variation $\delta N_a(z) = g_a(N(\mu + e\varphi(z), -E_a, \delta E_a) - N(\mu, -E_a, \delta E_a))$ and $\delta p(z) = g_p(P(\mu + e\varphi(z), -E_p, \delta E_p) - P(\mu, -E_p, \delta E_p))$ can be calculated from Eq.(2) valid in the static limit ($\omega=0$). Expressions for the functions $N$ and $P$ are given by Eqs.(3)-(5). Expressions for $\varphi(z)$ are given by Eqs. (17) in the linear Debye approximation for $\varphi$; Eqs. (20) include the nonlinearity up to $\varphi^3$.

In the static limit Eq.(30) could be simplified under the assumption of a weak field-induced bend bending as:

$$u_3(0) \approx -\sum_m w_m \int_0^h dz \varphi(z) =$$

$$= -\sum_m w_m R_S \begin{cases} \varphi_0(h,V) \cdot \left(1 - \exp\left(-\dfrac{h}{R_S}\right)\right)^2, & |eV/\mu| \ll 1 \\ -\sqrt{\dfrac{8}{\beta}} \left( \operatorname{arccoth}\left(\dfrac{3\sqrt{2\beta}\varphi_0(V)}{3\exp(h/R_S) - 2\alpha\varphi_0}\right) - \operatorname{arccoth}\left(\dfrac{3\sqrt{2\beta}\varphi_0}{3 - 2\alpha\varphi_0}\right)\right), & |eV/\mu| \sim 1. \end{cases} \quad (31)$$

The constants $w_a = g_a e \dfrac{dN(\mu, -E_a, \delta E_a)}{d\mu}\left(\beta_{33}^a - \dfrac{2s_{12}\beta_{11}^a}{s_{11} + s_{12}}\right)$ for acceptors and $w_p = 2g_p e \dfrac{dP(\mu, -E_p, \delta E_p)}{d\mu}\left(\beta_{33}^p - \dfrac{2s_{12}\beta_{11}^p}{s_{11} + s_{12}}\right)$ for holes. The first raw corresponds to the arbitrary film thickness, but the bend bending is regarded so small that expression for $\varphi_0(h,V)$ is given by Eq.(17b). The second raw is valid for stronger bend bending in the thicker films; for the case $\varphi_0(V)$ is determined from the Eq.(20b), where $V \approx U^* + U_b$.

The dependence of the static surface displacement on the acting voltage $V$ is shown in **Fig.13** for different values of film thickness $h$ and characteristics of acceptors band.

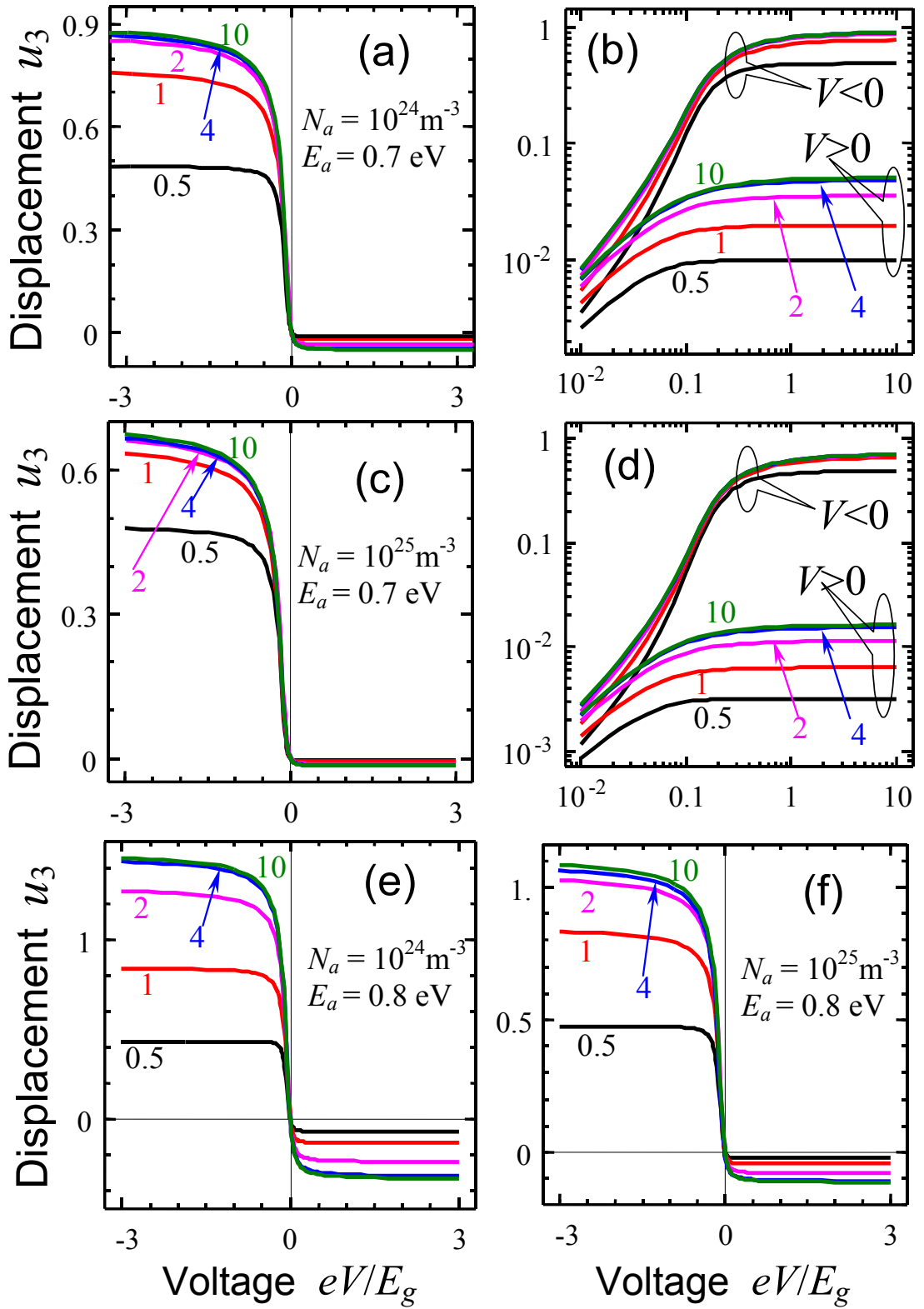

**Fig. 13.** Static surface displacement $u_3$ normalized on the value $N_a R_S (\beta_{33} - 2s_{12}\beta_{11}/(s_{11}+s_{12}))$ calculated for different film thickness $h/R_S$=0.5, 1, 2, 4, 10 (see numbers near the curves) and $E_a = 0.7$ eV (a, c), 0.8 eV (e, f), $N_a = 10^{24}$ m$^{-3}$, (a, e), $N_a = 10^{25}$ m$^{-3}$ (c, f), and $N_d = 10^{20}$ m$^{-3}$. Other parameters are the same as in **Fig. 4**. Plots (b, d) represent the same dependences as plots (a, c), but in double logarithmic scale. Calculated parameters: $R_S$ = 3.0, 1.2, 1.7, 0.64 nm and $\mu$ = −0.80, −0.85, −0.83, −0.88 eV for the plots a, b, c, d correspondingly.

It is seen for **Fig.13** that both applied voltage and film thickness increase lead to the strain response increase, being steep at small value and saturated for higher values of voltage and thickness, so that the response is almost constant for voltages much higher than $E_g$ and thickness much higher then $R_S$. It is consistent with the evolution of ionized acceptors distribution with voltage (see **Figs. 4**), since the changes in the distribution are essential only for small voltages and localized in the near-surface region with width of several $R_S$. The response dependence on the concentration of acceptors is nonlinear (compare plots (a, b) with (c, d)), despite the absolute value of the surface displacement increases ($u_3$ is roughly proportional to $N_a R_S$), the relative values decreases. The latter is related to the changes in Fermi level position, so that the concentration of ionized acceptors nonlinearly depends on $N_a$. For the small values $N_a$ Fermi level shifts to the conduction band and concentration of ionized acceptors decreases, so that the response drops to the very small values. Note, that normalized strain-voltage response is virtually independent on the screening radius $R_S$.

Asymmetry related with the voltage sign is explained by the strongly asymmetrical dependence of ionized acceptors distribution with voltage (see **Fig. 4**). The asymmetry strengthens with increase of $N_a$ (with respect to $N_d$) and with the shift of acceptors level in the direction of conduction band.

### 6.2. Linear frequency spectra of the strain-voltage response: Boltzmann approximation

Using the results obtained in the Section 5.2 and assuming that $\delta c(z,\omega) \approx \rho_\omega(z)/e$, we derived from Eq.(30) and **Table 2** the approximate analytical expression for the surface displacement of the ionic semiconductor film:

$$u_3(\omega) \approx -\left(\beta_{33} - \frac{s_{12}(\beta_{11}+\beta_{22})}{s_{11}+s_{12}}\right)\int_0^h dz \frac{\rho_\omega(z)}{e} = -\left(\beta_{33} - \frac{s_{12}(\beta_{11}+\beta_{22})}{s_{11}+s_{12}}\right)\frac{q_\omega}{e}$$

$$= -\frac{(\beta_{33} - s_{12}(\beta_{11}+\beta_{22})/(s_{11}+s_{12}))\varepsilon_0 \varepsilon_{33}^S \lambda (1-\exp(k(\omega)h))^2}{(1+\exp(2k(\omega)h))\left(i\omega\varepsilon_0\varepsilon_{33}^S + \lambda\frac{\tanh(hk(\omega))+k(\omega)\widetilde{H}}{(h+\widetilde{H})k(\omega)}\right)} \cdot \frac{V_0}{h+\widetilde{H}} \quad (32)$$

Here $\widetilde{H} = \frac{\varepsilon_{33}^S}{\varepsilon_{33}^g}H$, the spatial scale $k(\omega) = \sqrt{\frac{i\omega}{D}+\frac{1}{R_S^2}}$ and voltage $V(t) = V_0 \exp(i\omega t)$.

Note, that Eq.(32) corresponds to the linear response and so it is valid at $|eV/\mu| \ll 1$. Eq.(32) is derived for the case $J_\omega^c(0)=0$, $\rho_\omega(h)=0$. In the linear approximation other types of the boundary conditions (namely $J_\omega^c(0)=J_\omega^c(h)=0$ and $\rho_\omega(0)=\rho_\omega(h)=0$) lead to the total charge absence $q_\omega(\omega)=0$ and consequently zero surface displacement $u_3(\omega)=0$ (see **Table 2** and comments to it). Thus **Figs. 7c,d** also represents the frequency spectra of displacement (32), since $u_3(\omega) \sim q_\omega(\omega)$.

**Fig. 14** shows the parametric plot of the real part of the surface displacement $u_3$ vs. the real part of applied voltage calculated from Eq.(32). The dependences mimic hysteresis loops of purely ellipsoidal shape.

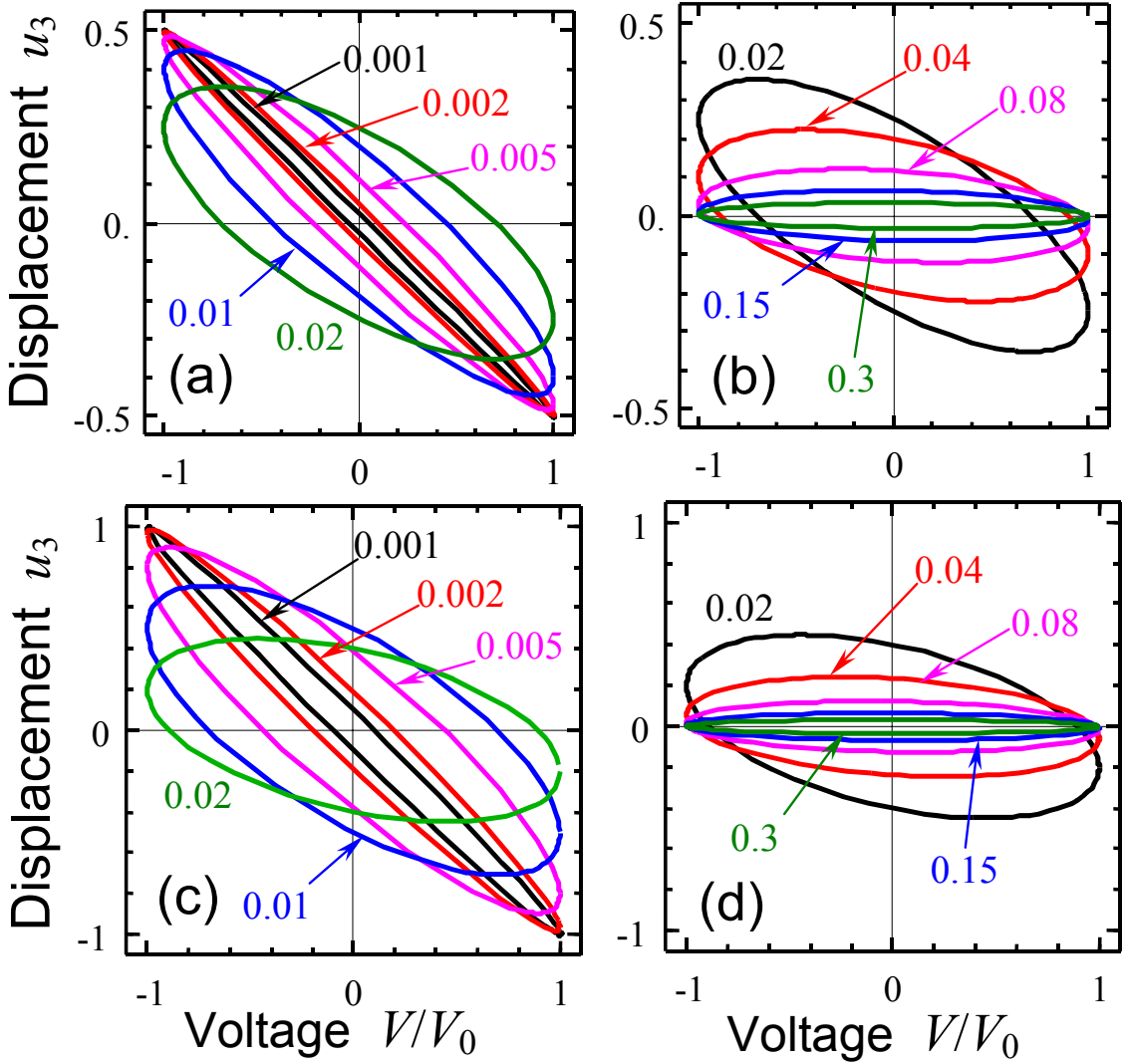

**Fig. 14.** Hysteresis loops: real part of the surface displacement $u_3$ vs. the real part of applied voltage for different values of dimensionless frequency $w\tau_M$ (see numbers near the curves) calculated for the gap thicknesses $\widetilde{H}/R_S = 0$ (a, b) and 1 (c, d). Film thickness $h/R_S = 100$, boundary conditions $J^c_\omega(0) = 0$, $\rho_\omega(h) = 0$. Plots (a, c) and (b, d) represent lower and higher frequencies respectively.

It is seen from **Fig. 14** that the frequency increase leads to the loops rotation, at the same time the loop inflation is maximal for intermediate frequencies, which is related to the maximum of the charge spectra imaginary part (see **Fig. 7c**). More complex shape of dynamic strain-voltage loops corresponds to the nonlinear response and will be analyzed in the next subsection.

### *6.3. Nonlinear strain–voltage response at finite frequences: calculations for rectangular DOS and in the Boltzmann approximation*

Below we analyze the dynamic strain-voltage response caused by the mobile ionized acceptors and holes in the ionic semiconductor film with negligible impact of the hopping conductivity. Note, that the dynamic strain-voltage response caused by the mobile ionized donors and electrons and corresponding currents can be analyzed in a similar way.

Since the imposed boundary conditions $J_c^a(0) = J_c^a(h) = 0$ are acceptor blocking and $\gamma_a(N_a^-, p) = 0$, the continuity equation rules that $\dfrac{d}{dt}\int_0^h N_a^-(z)dz = 0$. Thus, only the total changes of *the holes concentrations* contribute into the film surface displacement $u_3(V)$ via the hole-phonon deformation potential, i.e. $u_3(V) \sim \beta_{ij}^p \langle p(V) \rangle / N_p$. Acceptor contribution only add voltage independent constant base to the $u_3(V)$, and we omit the constant value below.

Since all nonzero components of the tensor $\beta_{ij}^p$ originate from the deformation potential of the hole-phonon coupling, which trace $\sum_i \beta_{ii}^p$ is proportional to the Fermi level $\mu$ in the metals and half-metals, corresponding *strain-voltage* response $u_3(V)$ can be unambiguously associated with the local response of *band structure* in the correlated oxides with high conductivity. Consequently the SPM measurements of the surface displacement in correlated oxide films could provide important information about their local band structure, once the tip electrode is ion blocking. Tensors $\beta_{ij}^p$ for some correlated oxides materials are listed in the **Table 1**.

SPM strain – voltage response calculated numerically for the external voltage frequency range $w\tau_M = 0.001 - 0.1$ is shown in **Figs. 15-16.** Different loops in **Figs. 15-16** corresponds to the increasing voltage amplitude $V_0$. Hereinafter we introduce the dimensionless strain $\tilde{u}_3 \sim \langle p \rangle / N_p$ and $w = \omega/2\pi$.

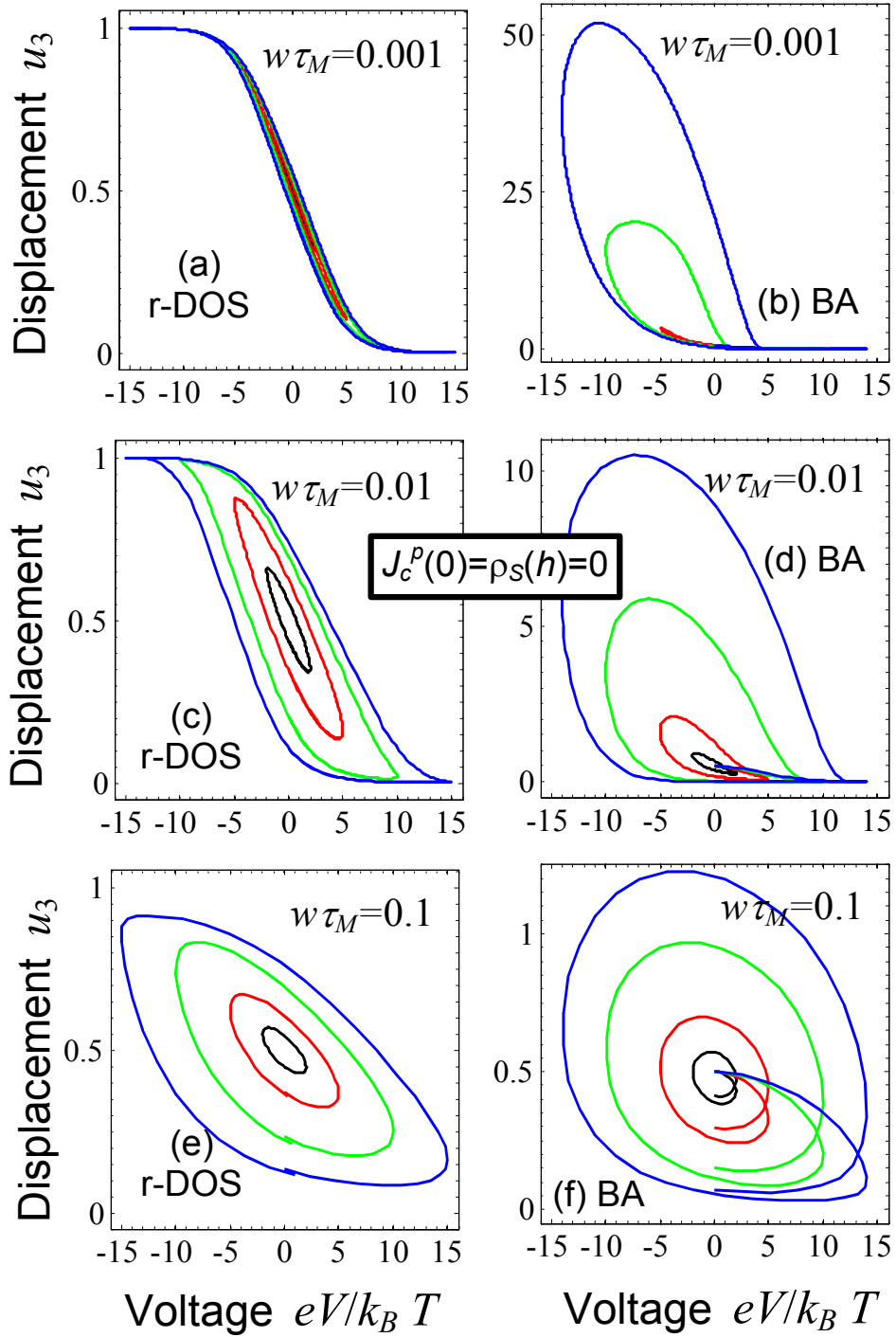

**Fig. 15.** SPM strain-voltage response calculated for different frequencies: $w\tau_M = 0.001$ (a, b), $w\tau_M = 0.01$ (c, d), and $w\tau_M = 0.1$ (e, f). Different loops (black, red, green and blue ones) correspond to the different values of maximal voltage $V_0 = 2, 5, 10, 15$ (in $k_B T/e$ units). Plots (a, c, e) are generated using rectangular DOS for the chemical potential (21) and plots (b, d, f) are generated in the Boltzmann approximation. DOS parameters $\delta E_a/k_B T = 2$, $\delta E_p/k_B T = 20$, $N_a = N_p$; film thickness $h/R_S = 5$, mobilities ratio $\eta_a/\eta_p = 0.1$ and $\gamma_a(N_a^-, p) = 0$. Asymmetric boundary conditions $\rho_S(h) = 0$, $J_c^a(h) = 0$ and $J_c^a(0) = J_c^p(0) = 0$ are imposed.

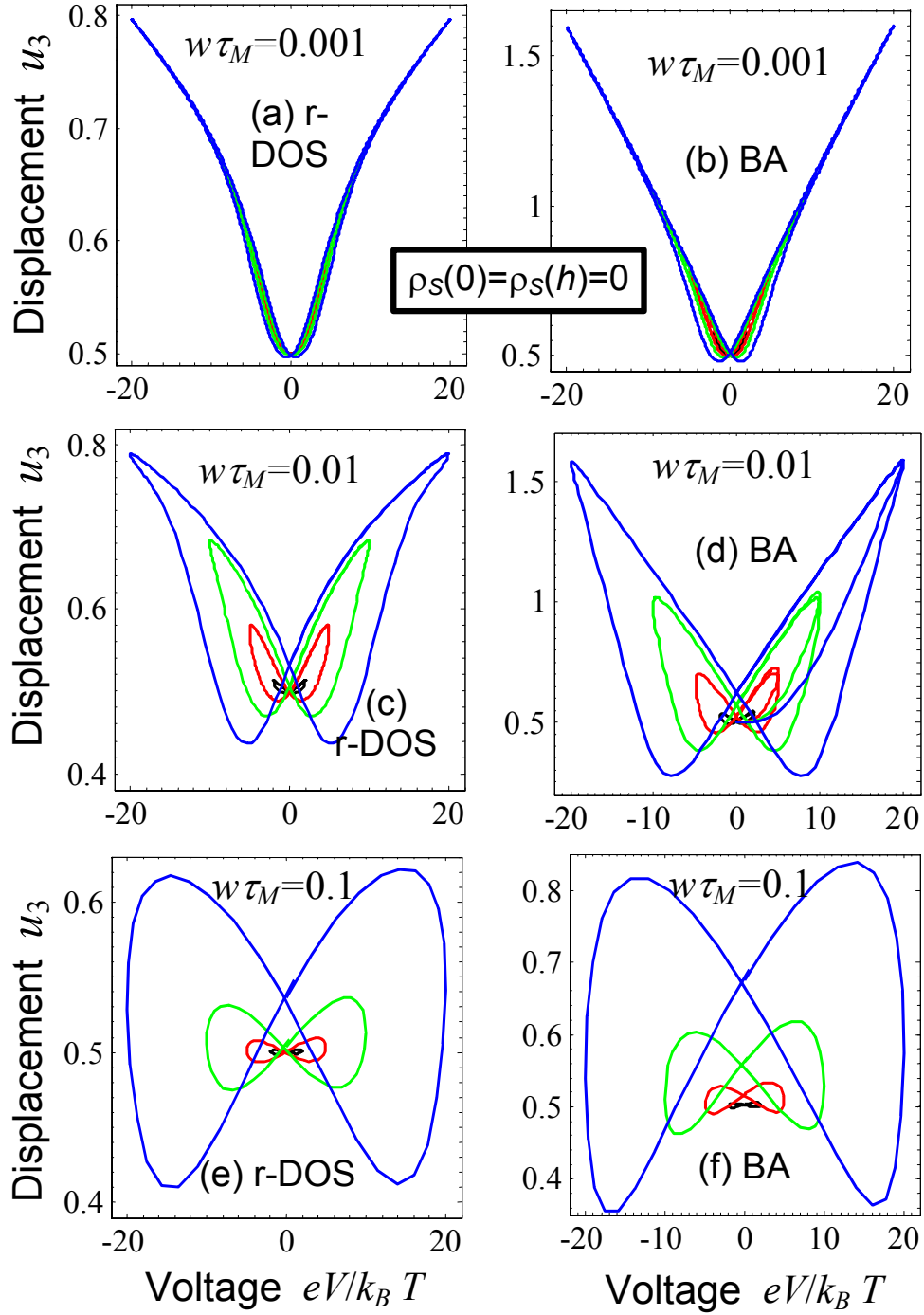

**Fig. 16.** SPM strain-voltage response calculated for different frequencies: $w\tau_M = 0.01$ (a, b), $w\tau_M = 0.03$ (c, d) and $w\tau_M = 0.1$ (e, f). Different loops (black, red, green and blue ones) correspond to the different values of maximal voltage $V_0 = 2, 5, 10, 20$ (in $k_BT/e$ units). Plots (a, c, e) are generated using rectangular DOS for the chemical potential (21) and plots (b, d, f) are generated in the Boltzmann approximation. DOS parameters $\delta E_a/k_BT = 2$, $\delta E_p/k_BT = 20$, $N_a = N_p$; film thickness $h/R_S = 5$, mobilities ratio $\eta_a/\eta_p = 0.1$ and $\gamma_a(N_a^-, p) = 0$. Symmetric boundary conditions $\rho_S(0) = \rho_S(h) = 0$ and $J_c^a(0) = J_c^a(h) = 0$ are imposed.

Numerical calculations, performed for the ion-blocking boundary conditions, $J_c^a(0) = J_c^a(h) = 0$, demonstrated that the nonlinear periodic displacement of the semiconductor film surface is proportional to the average concentration of holes, $\langle p \rangle = \frac{1}{h}\int_0^h p(z)dz$. In particular, periodic surface displacement $u_3(V)$ appears in response to the voltage $V(t) = V_0 \sin(\omega t)$ for the case of the asymmetric mixed-type boundary conditions, $J_c^p(0) = 0$, $\rho_S(h) = 0$ [shown in **Figs. 15**], as well as for symmetric "holes conducting" boundary conditions, $\rho_S(0) = \rho_S(h) = 0$ [shown in **Figs. 16**], despite its absence in the linear approximation. The nonlinear strain–voltage response is absent for the holes blocking conditions $J_c^p(0) = J_c^p(h) = 0$.

The strain-voltage loops, $u_3(V)$, shown **Figs. 15-16**, were calculated without hopping function ($\gamma_a(N_a^-, p) = 0$). They demonstrate principally different loop shapes, at that the differences mainly originate from the type of boundary conditions (compare with **Figs. 9-11**), namely:

(I) The hysteresis-like loops, shown in **Figs. 15**, are calculated for the case of asymmetric mixed-type boundary conditions, $J_c^p(0) = 0$, $\rho_S(h) = 0$. For the case the average density of holes and acceptors are time independent, despite the distributions changes in time. The loops calculated for rectangular DOS and in the Boltzmann approximation are strongly asymmetric with respect to the voltage sign both. Their shape is ellipsoidal only at small voltage amplitudes $V_0 < k_B T / e$ and becomes asymmetric hysteresis-like for rectangular DOS (or even irregular drop-like in the Boltzmann approximation) with $V_0$ increase. The loops calculated in the Boltzmann approximation are more irregular in comparison with the ones calculated for rectangular DOS. The loops becomes noticeably inflated with the frequency increase $w\tau_M \geq 0.1$, but the inflation is stronger in the Boltzmann approximation.

(II) The butterfly-like loops, shown in **Figs. 16**, are calculated for the case of symmetric holes conducting boundary conditions, $\rho_S(0) = \rho_S(h) = 0$. Note, that for the case the gaps should be absent. The loops generated at low frequencies $w\tau_M = 0.001$-$0.01$ are symmetric with respect to the voltage sign even after the first cycling. The loops generated at higher frequencies $w\tau_M = 0.1$ become symmetric with respect to the voltage sign only after relatively long relaxation of the initial conditions. The relaxation is more rapid for the loops calculated using rectangular DOS that for the ones calculated in Boltzmann approximation (compare plots c and d). The loops calculated in the Boltzmann approximation are more overblown and have pronounced linear parts in comparison with the ones calculated for rectangular DOS. The asymmetry $V \rightarrow -V$ disappears only for the symmetrical boundary conditions.

The discrepancies between current-voltage loops calculated from expressions Eqs.(21) using rectangular DOS for Fermi quasi-levels (plots a,c,e) and in the Boltzmann approximation (plots b,d,f) are essential both regarding the loop shape and the amplitude, at that the amplitude variation reaches the several orders of magnitude for the case of hole conducting boundary conditions. The discrepancy increases with the voltage amplitude $V_0$ increase and its frequency $w$ decrease. As it was argued in the Section 5, the Boltzmann approximation becomes invalid with $V_0$ increase, while the flattening and saturation of the carriers concentration naturally appear near the film interfaces, which is not accounted in the Boltzmann approximation (see **Figs. 6**).

Finally, let us analyze the dynamic strain-voltage response caused by the *immobile* acceptors and mobile holes allowing for *the local generation-recombination of holes* in the ionic semiconductor film. Typical voltage dependences of the film surface displacement are shown in **Figs. 17** for nonzero generation-recombination impact $\gamma_a(N_a^-, p) \neq 0$ and mixed boundary conditions $\rho_S(h) = 0$, $J_c^p(0) = 0$. For the mixed-type boundary conditions the contributions of ionized acceptors and holes into the strain response are essentially different: the holes contribution appeared strongly dependent on the voltage amplitude and frequency (compare **Figs.17** with **Figs.12e** and **Table 2**), the contribution of immobile ionized acceptors is almost independent on applied voltage and its frequency, its value is slightly deviates from the stationary concentration $N_a^- = \dfrac{2N_a}{1+\sqrt{1+4\gamma_R^a N_a / \gamma_G^a}}$. Other conditions lead to the voltage-independent displacement corresponding to the voltage independent stationary point of the system (22): $N_a^- = p = \dfrac{2N_a}{1+\sqrt{1+4\gamma_R^a N_a / \gamma_G^a}}$, as follows from the **Figs. 12d,f** and Eq.(30).

Note, that the "crosses" shown in **Figs. 17** are strongly different from the loops shown in **Figs. 15-16,** where $\eta_a \neq 0$ and $\gamma_a(N_a^-, p) = 0$. However, similarly to the strain-voltage of semiconductor film with mobile acceptors and without local generation-recombination acts, mobile holes mainly contribute into the strain-voltage response of the film with immobile acceptors and nonzero generation-recombination impact. The strength of the contribution is proportional to the deformation potential $\beta_{ij}^p$ originated from the electron-phonon coupling. The differences between the SPM strain-voltage response of mobile and immobile acceptors, as well as hopping conductivity impact, could be distinguished experimentally, at that the differences increases with the external voltage frequency decrease.

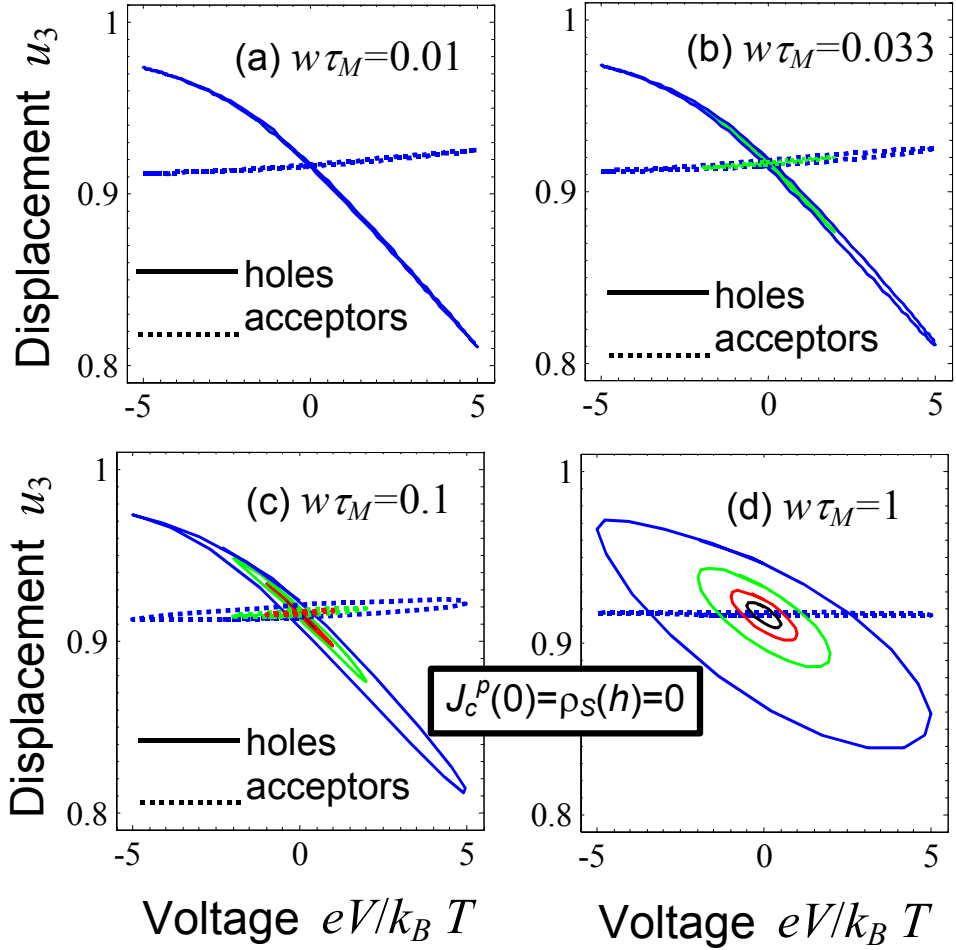

**Fig. 17.** Holes (solid curves) and ionized acceptors (dotted curves) contributions into the SPM strain-voltage response calculated for different frequencies of applied voltage: $w\tau_M = 0.01$ (a), 0.033 (b), 0.1 (c), 1 (d); acceptors mobility $\eta_a = 0$, $\tau_M \gamma_G^a = 0.3$, $\gamma_R^a N_a / \gamma_G^a = 0.1$. Different curves (black, red, green and blue) correspond to the different maximal voltage in $k_B T/e$ units. Rectangular DOS parameters $\delta E_a / k_B T = 1$, $\delta E_p / k_B T = 10$, $N_a = N_p$; film thickness $h/R_S = 5$. Asymmetric mixed-type boundary conditions $J_c^a(0) = J_c^p(0) = 0$ and $\rho_S(h) = 0$, $J_c^a(h) = 0$ are imposed.

To summarize the results of the Section 6, we evolve analytical formalism capable to describe the SPM strain-voltage response frequency spectra (linear approximation). Also we predict the great variety of the nonlinear static response and dynamic strain-voltage loops of the thin ionic semiconductor films with mobile acceptors and holes. Similarly to the current-voltage response, we demonstrated that conventional Boltzmann approximation becomes invalid for the description of the strain-voltage response with the increase of maximal voltage amplitude $V_0 > k_B T/e$, and the realistic DOS should be used for the correct calculations of the Fermi-quasi levels and generalized fluxes.

When the SPM tip and substrate electrode are acceptor blocking, the changes in holes concentration via the electron-phonon coupling mainly contribute into the film surface mechanical displacement strain, measured by strain SPM. The strength of the coupling is proportional to the deformation potential, the latter in turn may be stimulated by the local Jahn-Teller distortion existing in correlated oxides like $La_{1-x}Sr_xMnO_3$ and $La_{1-x}Sr_xCoO_3$. This allows us to relate the calculated strain-voltage response with the *local deformation potential* of correlated oxides. Consequently the SPM measurements of the local surface displacement with ion blocking tip electrode could provide important information about the local oxidation level, electron-phonon interactions via the deformation potential and even Jahn-Teller distortions in films of correlated oxides. Note, that the strain-voltage response of the film with mobile ionized donors and electrons can be analyzed in a similar way.

## Summary remarks

The paper is devoted to the analytical calculations of the electronic properties, current-voltage and local strain-voltage response in the heterostructure like "SPM tip electrode/dielectric gap/ionic semiconductor film /substrate electrode".

We derive analytical expressions for the dependences of the charge carriers density on the electrochemical potential and the band structure of the strongly doped ionic semiconductor film, assuming the stretched exponential DOS of the (mobile) donors, acceptors, electrons and holes. Then we derived analytical expressions for the static distributions of the electrostatic potential, field and charge carriers in the heterostructure. The analytical results were obtained in the Debye screening theory and bejound the linear approximation. Also we demonstrated that conventional Boltzmann approximation becomes invalid for the description of the current-voltage and strain-voltage response with the increase of maximal voltage amplitude, and the realistic DOS should be used for the correct calculations of the Fermi-quasi levels and generalized fluxes.

We performed analytical calculations of the strain-voltage response of the ionic semiconductor film caused by the local changes of (a) ions concentration (*stoichiometry contribution*); (b) acceptors (donors) charge state (*recharging contribution via ionic radius variation*); (c) free electrons (holes) concentration (*electron-phonon coupling via the deformation potential*). The contribution (b) into the strain-voltage SPM was not calculated previously, while the contribution (c) was not even predicted before, while our estimations performed for correlated oxides show that strength of (c) appeared comparable with (a,b).

When the SPM tip and substrate electrode are acceptor blocking, the changes in holes concentration via the electron-phonon coupling mainly contribute into the film surface mechanical displacement strain, measured by strain SPM. The strength of the coupling is proportional to the deformation potential, the latter in turn may be stimulated by the local Jahn-Teller distortion existing in correlated oxides like $La_{1-x}Sr_xMnO_3$ and $La_{1-x}Sr_xCoO_3$. This allows us to relate the calculated strain-voltage response with the *local deformation potential* of correlated oxides. Consequently the SPM measurements of the surface displacement with ion blocking tip electrode could provide important information about the local oxidation level, electron-phonon interactions via the deformation potential and even Jahn-Teller distortions in films of correlated oxides.

Moreover, using the independent measurements of voltage – capacitance characteristics (i.e. the dependence of the total electric charge on applied voltage), one could determine the coupling constants (*deformation potential*) between the strain and charge by comparison the voltage–charge with the voltage–strain characteristics. The seeming problem is the mixed contribution of different charged species as follows from Eq. (30). However, the problem could be overcome, since the ionic concentration is almost independent on the voltage for the case of the ion-blocking tip and electrodes. So, after eliminating the voltage independent acceptor parts from the measured capacitance and displacement, one could use the following relation between the *deformation potential* components $\beta_{ii}^p$ and measured displacement and concentration of non-blocked holes: $\dfrac{s_{12}(\beta_{11}^p + \beta_{22}^p)}{s_{11} + s_{12}} - \beta_{33}^p = \delta u_3(z=0,\omega)\left(\int_0^h dz\,\delta p(z,\omega)\right)^{-1}$. Note, that the strain-voltage response of the film with mobile ionized donors and electrons can be analyzed in a similar way.

Thus we evolve analytical formalism capable to describe the current-voltage and strain-voltage response frequency spectra in thin ionic semiconductor films. Also we predict the great variety of the nonlinear static and dynamic current-voltage and strain-voltage response of the film with mobile acceptors and holes. Note, that the response of the film with mobile ionized donors and electrons can be analyzed in a similar way. Calculated responses mimic hysteresis loops with pronounced memory window and double loops, observed experimentally in some ionic semiconductors (correlated oxides and resistive switching materials like *p*-$La_{1-x}Sr_xMnO_{3-\delta}$ and $La_{1-x}Sr_xCoO_{3-\delta}$). Predicted strain-voltage hysteresis of piezoelectric-like and butterfly-like shape requires experimental justification by SPM.

**Acknowledgements**

Research sponsored by Ministry of Science and Education of Ukrainian and National Science Foundation (Materials World Network, DMR-0908718). EAE and ANM gratefully acknowledge financial support from National Academy of Science of Ukraine The research is supported in part (SVK) by the Division of Scientific User Facilities, DOE BES. EAE and ANM gratefully acknowledge multiple discussions with Prof. N.V. Morozovskii.

## Appendix A. Notes to the static solution

Note to Eq.(14):

Thus Eq.(8b) along with the boundary conditions at $z=0$ and $z=h$ acquires the form:

$$\begin{cases} \dfrac{d^2\varphi}{dz^2} = -\dfrac{e}{\varepsilon_{33}^S \varepsilon_0} \left( \begin{array}{l} \int\limits_{-\infty}^{\infty} \left(g_p(\varepsilon)f(\mu+e\varphi-\varepsilon) - g_n(\varepsilon)f(\varepsilon-\mu-e\varphi)\right)\cdot d\varepsilon \\ + \int\limits_{-\infty}^{\infty} d\varepsilon \cdot \left(g_d(\varepsilon)f(e\varphi+\mu-\varepsilon) - g_a(\varepsilon)f(\varepsilon-e\varphi-\mu)\right) \end{array} \right), \\ \varphi(0) - H \dfrac{\varepsilon_{33}^S}{\varepsilon_{33}^g} \dfrac{\partial \varphi(0)}{\partial z} = U^* + U_b + \dfrac{H\sigma_f}{\varepsilon_0 \varepsilon_{33}^g}, \quad \varphi(h)=0. \end{cases} \quad (A.1)$$

The functional, which minimization leads to Eqs.(A.1a), was derived as:

$$F = \int\limits_0^h dz \left( \dfrac{1}{2}\left(\dfrac{d\varphi}{dz}\right)^2 + \dfrac{e}{\varepsilon_{33}^S \varepsilon_0} \left( \begin{array}{l} \int\limits_{-\infty}^{\infty} d\varepsilon \int\limits_0^{\varphi} d\phi \left(g_p(\varepsilon)f(\mu+e\phi-\varepsilon) - g_n(\varepsilon)f(\varepsilon-\mu-e\phi)\right) \\ + \int\limits_{-\infty}^{\infty} d\varepsilon \int\limits_0^{\varphi} d\phi \left(g_d(\varepsilon)f(\mu+e\phi-\varepsilon) - g_a(\varepsilon)f(\varepsilon-\mu-e\phi)\right) \end{array} \right) \right)$$
$$+ \dfrac{\varepsilon_0 \varepsilon_{33}^g}{2H}\left(\varphi(0) - U^* - U_b - \dfrac{H\sigma_f}{\varepsilon_0 \varepsilon_{33}^g}\right)^2 \quad (A.2)$$

Eq.(A.2) should be minimized along with the boundary condition $\varphi(h)=0$.

Note to Eqs.(19):

the first integral

$$\left(\dfrac{d\varphi}{dz}\right)^2 - \dfrac{\varphi^2}{R_S^2}\left(1 + \dfrac{2\alpha}{3}\varphi + \dfrac{\beta}{2}\varphi^2\right) = 0, \quad \rightarrow \quad \dfrac{\pm d\varphi}{\varphi\sqrt{1+(2\alpha/3)\varphi+(\beta/2)\varphi^2}} = \dfrac{dz}{R_S}$$

the second integral $\ln\left(\dfrac{\varphi}{\varphi_0}\right) - \ln\left(2 + \dfrac{2\alpha}{3}\varphi + 2\sqrt{1+\dfrac{2\alpha}{3}\varphi+\dfrac{\beta}{2}\varphi^2}\right) = \pm\dfrac{z}{R_S}$  (A.3)

Note to (20b):

$$\varphi_0 = \dfrac{V^*}{4\left(3+V^*\alpha \pm \sqrt{9+6V^*\alpha+\dfrac{9}{2}(V^*)^2\beta}\right)}, \quad V^* \equiv V\left(1+\dfrac{H}{R_S}\dfrac{\varepsilon_{33}^S}{\varepsilon_{33}^g}+\ldots\right)^{-1} \quad (A.4)$$

The sign in the denominator (20b) should be determined from the condition $\alpha\varphi_0 < 3/2$.

The necessary condition of bistability appearance at zero gap $H=0$ can be reduced to the appearance of the multiple roots (besides the trivial one $V=0$) of the transcendental equation

$$\begin{pmatrix} 2g_n N(\mu+eV, E_n, \delta E_n) - g_d P(\mu+eV, -E_d, \delta E_d) \\ + g_a N(\mu+eV, -E_a, \delta E_a) - 2g_p P(\mu+eV, -E_p, \delta E_p) \end{pmatrix} = 0 \quad \text{(A.5)}$$

e.g. by the graphical method in the space of parameters $\{E_m, \delta E_m, g_m\}$, $m = a, d, n, p$. Unfortunately, no bistability was detected.

## Appendix B. System dynamic response

In the external field changing with arbitrary frequency the continuity equation $\frac{\partial \rho}{\partial t} + \text{div} J_c = 0$ should be valid, which along with the electrodynamics equations $\text{div} E_e = \frac{\rho}{\varepsilon_0 \varepsilon}$, $\text{div} B = 0$, $\text{rot} E_e = -\frac{\partial B}{\partial t}$ and $\text{rot} B = \mu_0 \left( J_c + \varepsilon_0 \varepsilon \frac{\partial E_e}{\partial t} \right)$ leads to: $\text{rot} \frac{\partial B}{\partial t} = \mu_0 \frac{\partial}{\partial t} \left( J_c + \varepsilon_0 \varepsilon \frac{\partial E_e}{\partial t} \right)$. So that

$$-\text{rot}(\text{rot} E_e) = \mu_0 \frac{\partial}{\partial t}\left( J_c + \varepsilon_0 \varepsilon \frac{\partial E_e}{\partial t} \right) \text{ and}$$

$$-\frac{\partial}{\partial t}\mu_0\left( J_c + \varepsilon_0 \varepsilon \frac{\partial E_e}{\partial t} \right) = \text{rot}(\text{rot} E_e) = \text{grad}(\text{div} E_e) - \Delta E_e = \frac{\text{grad}(\rho)}{\varepsilon_0 \varepsilon} - \Delta E_e \quad \text{(B.1)}$$

Dimensionless system for a strongly doped semiconductor has a form:

$$\frac{d^2 \tilde{\varphi}}{d\tilde{z}^2} = -(\tilde{p} - \tilde{a}), \quad \text{(B.2a)}$$

$$-\frac{\partial \tilde{a}}{\partial \tilde{t}} + \frac{\eta_a}{\eta_p}\frac{\partial}{\partial \tilde{z}}\left( \tilde{a}\frac{\partial(\tilde{\zeta}_a - \tilde{\varphi})}{\partial \tilde{z}} \right) = R\tilde{a}\tilde{p} - G(1-\tilde{a})\theta(1-\tilde{a}), \quad \text{(B.2b)}$$

$$\frac{\partial \tilde{p}}{\partial \tilde{t}} + \frac{\partial}{\partial \tilde{z}}\left( \tilde{p}\frac{\partial(\tilde{\zeta}_p - \tilde{\varphi})}{\partial \tilde{z}} \right) = -R\tilde{a}\tilde{p} + G(1-\tilde{a})\theta(1-\tilde{a}), \quad \text{(B.2c)}$$

dimensionless variables, coordinate, time and constants:

$$\tilde{\varphi} = \frac{e\varphi}{k_B T}, \quad \tilde{z} = \frac{z}{R_S}, \quad \frac{1}{R_S^2} = \frac{e^2 N_p}{\varepsilon_{33}^S \varepsilon_0 k_B T}, \quad \tilde{a} = \frac{N_a^-}{N_p}, \quad \tilde{p} = \frac{p}{N_p},$$

$$\tilde{t} = \frac{t}{\tau_M}, \quad \tau_M = \frac{R_S^2 e}{\eta_p k_B T} \equiv \frac{\varepsilon_{33}^S \varepsilon_0}{\eta_p N_p e}, \quad \tilde{\zeta}_a = \frac{\zeta_a}{k_B T}, \quad \tilde{\zeta}_p = \frac{\zeta_p}{k_B T}, \quad R = \tau_M \gamma_R^a N_a, \quad G = \tau_M \gamma_G^a,$$

$$\delta \tilde{E}_a = \frac{\delta E_a}{k_B T}, \quad \tilde{E}_a = \frac{E_a}{k_B T}, \quad \delta \tilde{E}_p = \frac{\delta E_p}{k_B T}, \quad \tilde{E}_p = \frac{E_p}{k_B T}.$$

The stationary point of the system (B.2) is

$$\tilde{a} = \tilde{p} = \frac{2G}{G+\sqrt{G^2+4GR}} = \frac{2}{1+\sqrt{1+4R/G}} = \frac{2}{1+\sqrt{1+4\gamma_R^a N_a/\gamma_G^a}}.$$

## Appendix C. Dynamic response in the local charge density approximation

For the case the total electric current is the semiconductor film is $J = J_d + J_c$, where $J_d(z,t) = -\varepsilon_0 \varepsilon_{33}^{S,g} \frac{\partial E_z}{\partial t}$ is the displacement current (also existing in the gap), and $J_c(z,t) = \sum_m J_m = J_0(t) - \frac{\partial}{\partial t}\int_0^z \rho_S(z,t)dz$ is the full conductivity current (existing in the semiconductor only), which is in agreement with continuity equation $\frac{\partial \rho_S}{\partial t} + \frac{\partial J_c}{\partial z} = 0$.

The continuity equation should be solved along with the electrodynamics equations. In Appendix A we derived that it leads to the one equation for electric field inside the semiconductor film: $\frac{\partial^2 E_z}{\partial z^2} = \frac{1}{\varepsilon_0 \varepsilon_{33}^S}\frac{\partial \rho_S}{\partial z} + \left(\frac{\partial^2}{\partial t^2}\frac{\varepsilon_{33}^S}{c^2}E_z + \mu_0\frac{\partial J_0(t)}{\partial t} - \mu_0\frac{\partial^2}{\partial t^2}\int_0^z \rho_S(z,t)dz\right)$, where $\mu_0$ is the universal magnetic constant. Then we should solve the steady-state boundary problem for the periodic part of potential $\varphi(z,t) = \varphi_S(z) + \varphi_\omega(z)\exp(i\omega t)$, the space charge density variation as $\rho_S(z,t) = \rho_S(z) + \rho_\omega(z)\exp(i\omega t)$ and $\sigma_f(t) = \sigma_S + \sigma_\omega \exp(i\omega t)$.

Equation inside the semiconductor:

$$\frac{d^3\varphi_\omega}{dz^3} + \frac{\varepsilon_{33}^S \omega^2}{c^2}\frac{d\varphi_\omega}{dz} = -\frac{1}{\varepsilon_0 \varepsilon_{33}^S}\frac{d\rho_\omega}{dz} - i\omega\mu_0 J_\omega - \omega^2\mu_0\int_0^z \rho_\omega(x)dx$$
$$\approx \frac{1}{R_S^2}\left(\frac{d\varphi_\omega}{dz} + \varepsilon_{33}^S \frac{\omega^2}{c^2}\int_0^z \varphi_\omega(x)dx\right) - i\omega\mu_0 J_\omega \qquad (0 < z < h) \qquad (C.1a)$$

Note, that in Eqs.(C.1a) we expanded the space charge density variation as $\rho_\omega(z) = -\varepsilon_0 \varepsilon_{33}^S \frac{\varphi_\omega(z)}{R_S^2}$ assuming *the local density approximation validity*, e.g. the linear Debye approximation valid at $|e\varphi/\mu| \ll 1$. Equation inside the gap has the form:

$$\frac{d^3\varphi_\omega}{dz^3} + \frac{\varepsilon_{33}^g \omega^2}{c^2}\frac{d\varphi_\omega}{dz} = 0 \qquad (-H < z < 0) \qquad (C.1b)$$

Eqs.(C.1) should be supplemented with the boundary conditions for the potential and currents at the semiconductor/gap interface:

$$\varphi_\omega(-H) = u_0, \qquad \text{(metal tip-dielectric gap)} \qquad (C.2a)$$

$$\varphi_\omega(-0) - \varphi_\omega(+0) \approx 0, \quad \text{(semiconductor-dielectric gap)} \qquad (C.2b)$$

$$D_{2n} - D_{1n} = \sigma_\omega \quad \Rightarrow \quad -\varepsilon_0\left(\varepsilon_{33}^S \frac{d\varphi_\omega(+0)}{dz} - \varepsilon_{33}^g \frac{d\varphi_\omega(-0)}{dz}\right) = \sigma_\omega. \qquad (C.2c)$$

$$\frac{\partial D_{2n}}{\partial t} + J_{2n}^c = \frac{\partial D_{1n}}{\partial t} + J_{1n}^c \quad \Rightarrow \quad i\omega\varepsilon_0\varepsilon_{33}^S \frac{d\varphi_\omega(+0)}{dz} + J_\omega = i\omega\varepsilon_0\varepsilon_{33}^g \frac{d\varphi_\omega(-0)}{dz} \qquad (C.2d)$$

and the condition of the potential vanishing at the bottom electrode $\varphi_\omega(h) = 0$.

The compatibility of Eq.(C.2c) with (C.2d) leads to the condition $J_\omega = i\omega\sigma_\omega$. Potential distribution inside the gap $\varphi(z) = u_0 + \frac{H+z}{\varepsilon_0\varepsilon_{33}^g}\left(\sigma_\omega + \varepsilon_0\varepsilon_{33}^S \frac{\partial\varphi(+0)}{\partial z}\right)$ was obtained neglecting the terms proportional to the ratio $\omega^2/c^2$, which is obviously valid in the working frequency range $\omega^2 \ll \left(c^2/\varepsilon_{33}^g R_S^2\right) \sim 10^{32}$ 1/s2 without any significant loss of precision [16].

Then, similarly to Eq.(4-5), Eqs.(C.1) and Eqs.(C.2a-c) reduces to the boundary problem:

$$\begin{cases} \dfrac{d^3\varphi_\omega}{dz^3} + \dfrac{\varepsilon_{33}^S \omega^2}{c^2}\dfrac{d\varphi_\omega}{dz} \approx \dfrac{1}{R_S^2}\left(\dfrac{d\varphi_\omega}{dz} + \varepsilon_{33}^S \dfrac{\omega^2}{c^2}\int_0^z \varphi_\omega(x)dx\right) - i\omega\mu_0 J_\omega, \\ \varphi_\omega(0) - H\dfrac{\varepsilon_{33}^S}{\varepsilon_{33}^g}\dfrac{d\varphi_\omega(0)}{dz} = u_0 + \dfrac{H}{\varepsilon_0\varepsilon_{33}^g}\dfrac{J_\omega}{i\omega}, \quad \varphi_\omega(h) = 0. \end{cases} \qquad (C.3)$$

The solution of Eq.(C.3) has the form: $\varphi_\omega(z) = \sum_i C_i \exp(k_i z)$ and the imbalance condition $J_\omega = \varepsilon_0\varepsilon_{33}^S \dfrac{i\omega}{R_S^2}\sum_i \dfrac{C_i}{k_i}$ should be imposed. Characteristic equation $k^3 + \left(\dfrac{\varepsilon_{33}^S\omega^2}{c^2} - \dfrac{1}{R_S^2}\right)k - \dfrac{\varepsilon_{33}^S\omega^2}{c^2 R_S^2}\dfrac{1}{k} = 0$ of Eq.(C.3) has four roots $k_{1,2} = \pm\dfrac{1}{R_S}$, $k_{3,4} = \pm i\sqrt{\varepsilon_{33}^S\dfrac{\omega^2}{c^2}}$. In all subsequent calculations we neglect the terms proportional to the ratio $\dfrac{\omega^2}{c^2}$, which is obviously valid in the working frequency range $\omega^2 \ll \left(c^2/\varepsilon_{33}^S R_S^2\right) \sim 10^{32}$ 1/s2. Then two constants C1,2 corresponding to the roots k1,2 and Jω should be found from the system of linear equations:

$$\sum_i C_i\left(1 - H\dfrac{\varepsilon_{33}^S}{\varepsilon_{33}^g}k_i\right) - \dfrac{H}{\varepsilon_0\varepsilon_{33}^g}\dfrac{J_\omega}{i\omega} = u_0, \quad \sum_i C_i \exp(k_i h) = 0, \quad \dfrac{\varepsilon_0\varepsilon_{33}^S}{R_S^2}\sum_i \dfrac{C_i}{k_i} - \dfrac{J_\omega}{i\omega} = 0. \qquad (C.4)$$

Z-distributions of the potential and electric current acquire the form

$$\varphi(z) = \frac{u_0\left(\exp\left(-\frac{z}{R_S}\right) - \exp\left(\frac{z-2h}{R_S}\right)\right)R_S}{\frac{\varepsilon_{33}^S H}{\varepsilon_{33}^g}\left(1+\exp\left(-\frac{2h}{R_S}\right)\right) + 2\frac{\varepsilon_{33}^g H}{\varepsilon_{33}^S}\exp\left(-\frac{h}{R_S}\right) + R_S\left(1-\exp\left(-\frac{2h}{R_S}\right)\right)}, \quad 0 < z < h$$
(C.5a)

$$\varphi(z) = u_0 - \frac{u_0(z+H)\frac{\varepsilon_{33}^S}{\varepsilon_{33}^g}\left(1+\exp\left(\frac{-2h}{R_S}\right)\right) + 2\frac{\varepsilon_{33}^g}{\varepsilon_{33}^S}\exp\left(\frac{-h}{R_S}\right)}{\frac{\varepsilon_{33}^S H}{\varepsilon_{33}^g}\left(1+\exp\left(-\frac{2h}{R_S}\right)\right) + 2\frac{\varepsilon_{33}^g H}{\varepsilon_{33}^S}\exp\left(-\frac{h}{R_S}\right) + R_S\left(1-\exp\left(-\frac{2h}{R_S}\right)\right)}, \quad -H < z < 0$$

(C.5b)

$$J(z,\omega) = \begin{cases} -i\omega\left(\varepsilon_0\varepsilon_{33}^S\frac{d\varphi(z)}{dz} - \varepsilon_0\varepsilon_{33}^S\int_0^z \frac{\varphi_\omega(x)}{R_S^2}dx\right), & 0 \leq z \leq h, \\ -i\omega\varepsilon_0\varepsilon_{33}^S\frac{d\varphi(z)}{dz}, & -H \leq z \leq 0. \end{cases}$$
(C.5c)

Note, that the tunneling current in the gap is regarded negligibly small in comparison with the displacement current.

It is seen from the Eqs.(C.5) that the conventional local charge density approximation $\rho_\omega(z) = -\varepsilon_0\varepsilon_{33}^S\frac{\varphi_\omega(z)}{R_S^2}$ leads to the independence of the potential distribution on the external field frequency. The current $J(z,\omega)$ is simply proportional to the product $i\omega$, which corresponds to the pure capacitive reactance. The result looks unrealistic for existing semiconductors especially at high frequencies. So at least for the high-frequency case the non-locality of the current response to external periodic voltage should be considered.

Note, that the distributions of the different species concentrations (e.g. electron and acceptors) are proportional to the electric potential distribution in the linear approximation.